\documentclass[aps,prb,showpacs,amsmath,amssymb,twocolumn,superscriptaddress,letterpaper,english]{revtex4}
\usepackage{times}
\usepackage{amsfonts}
\usepackage{mathrsfs}
\usepackage{graphicx}
\usepackage{dcolumn}
\usepackage{bm}
\usepackage{color}

\usepackage[colorlinks,bookmarks=false,citecolor=blue,linkcolor=red,urlcolor=blue]{hyperref}
\usepackage{url}

\def\be{\begin{equation}}       \def\ee{\end{equation}}
\def\bea{\begin{eqnarray}}      \def\eea{\end{eqnarray}}

\newcommand{\eph}{$e$-$ph$}
\newcommand{\Tc}{$T_{\rm c}$}

\begin{document}
\title{Origin of the pressure-dependent {\Tc} valley in
  superconducting simple cubic phosphorus}

\author{Xianxin~Wu{$^*$}}
\affiliation{Institut f\"ur Theoretische Physik und Astrophysik,
  Julius-Maximilians-Universit\"at W\"urzburg, 97074 W\"urzburg,
  Germany}
\email{Xianxin.Wu@physik.uni-wuerzburg.de}

\author{Harald~O.~Jeschke}
\affiliation{Institut f\"ur Theoretische Physik und Astrophysik,
  Julius-Maximilians-Universit\"at W\"urzburg, 97074 W\"urzburg,
  Germany}
\affiliation{Research Institute for Interdisciplinary Science, Okayama
  University, Okayama 700-8530, Japan}

\author{Domenico~Di~Sante}
\affiliation{Institut f\"ur Theoretische Physik und Astrophysik,
  Julius-Maximilians-Universit\"at W\"urzburg, 97074 W\"urzburg,
  Germany}

\author{Fabian~O.~von~Rohr }
\affiliation{Department of Chemistry, Princeton University, Princeton, New Jersey 08544, USA }
\affiliation{Department of Chemistry, University of Zurich, CH-8057, Switzerland }

\author{Robert~J.~Cava}
\affiliation{Department of Chemistry, Princeton University, Princeton, New Jersey 08544, USA }

\author{Ronny~Thomale}
\affiliation{Institut f\"ur Theoretische Physik und Astrophysik,
  Julius-Maximilians-Universit\"at W\"urzburg, 97074 W\"urzburg,
  Germany}

\date{\today}

\begin{abstract}
  Motivated by recent experiments, we investigate the
  pressure-dependent electronic structure and electron-phonon ({\eph})
  coupling for simple cubic phosphorus by performing first-principle
  calculations within the full potential linearized augmented plane
  wave method. As a function of increasing pressure, our calculations
  show a valley feature in {\Tc}, followed by an eventual decrease for
  higher pressures. We demonstrate that this {\Tc} valley at low
  pressures is due to two nearby Lifshitz transitions, as we analyze
  the band-resolved contributions to the {\eph} coupling. Below the first
  Lifshitz transition, the phonon hardening and shrinking of the
  $\gamma$ Fermi surface with $s$ orbital character results in a
  decreased {\Tc} with increasing pressure. After the second Lifshitz
  transition, the appearance of $\delta$ Fermi surfaces with $3d$
  orbital character generate strong {\eph} inter-band couplings in
  $\alpha\delta$ and $\beta\delta$ channels, and hence lead to an
  increase of {\Tc}. For higher pressures, the phonon hardening
  finally dominates, and {\Tc} decreases again. Our study reveals that
  the intriguing {\Tc} valley discovered in experiment can be
  attributed to Lifshitz transitions, while the plateau of {\Tc}
  detected at intermediate pressures appears to be beyond the scope of
  our analysis. This strongly suggests that besides {\eph} coupling,
  electronic correlations along with plasmonic contributions may be
  relevant for simple cubic phosphorous. Our findings hint at the
  notion that increasing pressure can shift the low-energy orbital
  weight towards $d$ character, and as such even trigger an enhanced
  importance of orbital-selective electronic correlations despite an
  increase of the overall bandwidth.
\end{abstract}


\maketitle

\section{Introduction}

Layered black phosphorus (black P), a narrow-gap semiconductor, crystallizes in the
orthorhombic $A17$ structure at ambient pressure and temperature, with
layers weakly bound by van der Waals
forces\cite{Jamieson1963,Kikegawa1983}.  The monolayer phosphorene is
similar to graphene, albeit with a natural band
gap\cite{Reich2014,Liu2014,Li2014}. Recently, topologically
non-trivial semi-metallic properties have been discovered in the $A17$
structure at low pressure ({\it i.e.} a few
GPa)\cite{Xiang2015,Zhao2016,Gong2016,Li2016}, which contributed to
revitalizing the interest in this material.  Under
pressure, early studies showed that black phosphorus
exhibits several phase
transitions\cite{Jamieson1963,Kikegawa1983}. The structure changes
from $A17$ (orthorhombic) to semimetallic $A7$ (rhombohedral) at
4.5~GPa, followed by a metallic primitive cubic phase at
10~GPa\cite{Jamieson1963,Kikegawa1983}. The former transition is
sluggish while the latter is rather sharp. It is remarkable to note
that although the atoms are not closely packed~\cite{seo199926}, the primitive cubic (or simple cubic)
phase is stable up to a pressure of 107~GPa before it starts to
transform into a hexagonal phase, completed at
137~GPa\cite{Akahama1999}. Aside from the unusually rich variety of structural
phase transitions, superconductivity was discovered in the
rhombohedral and cubic phases above
5~GPa\cite{Wittig1985,Kawamura1984,Kawamura1985}. Put together, this
naturally suggests black P to be a prime candidate for a particularly
interesting shape of {\Tc} as a function of different trajectories
taken with respect to pressure and
temperature. Accordingly, the pressure dependence of {\Tc} has been
investigated by many groups. Still, there is no experimental consensus
on this question. This is because (i) the measured {\Tc} appears to be
strongly dependent on the path taken in $P$-$T$ parameter
space\cite{Kawamura1984,Kawamura1985} and (ii), probably not
uncorrelated with (i), there are mixed states due to incomplete
structural phase transformations. Kawamura \emph{et
  al.}\cite{Kawamura1984,Kawamura1985} found that {\Tc} increased
continuously with pressure in the cubic phase on one path, while it
varied only weakly with pressure around 6 K on the other path. Wittig
\emph{et al.}\cite{Wittig1985} discovered that {\Tc} showed two
distinct peaks at about 12 GPa and 23 GPa separated by an intriguing
valley centered at about 17 GPa. In particular, Karuzawa \emph{et
  al.}\cite{Karuzawa2002} showed that {\Tc} exhibits only a single
peak around 23 GPa, and the corresponding {\Tc} was about 9.5 K.  In
the most recent experiment, performed with high quality black P
crystals and excellent pressure cell conditions, Guo \emph{et
  al.}\cite{Guo2016} found a behavior similar to that of Wittig
\emph{et al.} in terms of pressure variation of {\Tc}: in the cubic phase,
with increasing pressure, {\Tc} first decreases and then increases,
i.e. it forms a {\Tc} valley. From about 32~GPa, {\Tc} exhibits a
plateau for an extended pressure regime at a comparably high
value. Moreover, Hall effect measurements suggested that hole carriers
play an important role in promoting superconductivity, and that the
{\Tc} valley may originate from a Lifshitz transition\cite{Guo2016}.

From an initial theoretical point of view, black P does not promise to
be particularly controversial. As the {\Tc} for cubic phosphorus is
relatively low (with a maximum of about 10~K) and phosphorous has an
approximately half-filled $p$ band, superconductivity has so far been
assumed to be driven by electron phonon ({\eph}) coupling. No
consensus, however, has been reached on the pressure dependence of
{\Tc} in the cubic phase, as the result strongly depends on the
adopted theoretical methods. By relating average phonon frequency and bulk
modulus, early calculations\cite{Rajagopalan1989} show a single peaked
structure for {\Tc} as a function of pressure, where the peak position is
nearby the second peak of the experimental data in
Ref.~\onlinecite{Wittig1985}. It must be noted, however, that the variation of the phonon
spectrum under pressure is completely omitted in this analysis. Other
calculations have mainly focused on the electronic
structure\cite{Aoki1987}, or find that {\Tc} changes little with
increasing pressure\cite{Nagara2010}, which is only consistent with
the data on one particular path through $P$-$T$
space\cite{Kawamura1984}; it is inconsistent with recent experimental
evidence. Another study finds that {\Tc} increases slightly from 8.5
to 11~K as the pressure increases from 10 to 35 GPa\cite{Nixon2010},
which again hardly bears similarity to experiments. A more recent, and
technically significantly advanced, study involving full \emph{ab
  initio} calculations of {\eph} coupling using pseudopotentials finds
that {\Tc} decreases monotonously in the simple cubic
phase\cite{Cohen2013}, and reports no indication for any valley or
plateau formation. Furthermore, in the latest theoretical
  studies, metastable structures in black P are found to be important to
explain the $P$-$T$ path dependent $T_c$ in
experiments\cite{Flores-Livas2017}.

In this paper, we pursue two objectives. First, facing partly
contradictory experimental evidence, we attempt to consider all experimental facts
on the same footing, but if in doubt center our experimental reference
around Ref.~\onlinecite{Guo2016}, and apply {\it ab initio} methods to
the most refined level in order to provide a theoretical
explanation. In particular, as justified by the crystal analysis
performed in Ref.~\onlinecite{Guo2016}, we explicitly start from the hypothesis that the
structural transition to the simple cubic phase sets in comparably
early, and that the {\Tc} valley, as observed, occurs within this phase. By
employing more reliable full-potential calculations instead of
pseudopotential calculations, we attempt to resolve previous
discrepancies in theoretical calculations. We manage to explain the
{\Tc} valley formation found experimentally, as the precision of our
analysis allows us to disentangle effects of Lifshitz transitions from
structural transitions. Second, we try to sharpen the perspective on
how the {\it ab initio} analysis would have to be extended in order to
refine the correspondence between theory and experiment. As we
investigate electronic correlations in simple cubic black P, we notice several
interesting aspects that might be relevant for a larger class of
materials under pressure. For instance, we find that the $d$-orbital
low energy weight increases as a function of pressure. In total, we
are led to conclude that orbital-selective interactions as well as plasmonic
contributions might provide central additional insights into the persistent
intricacies of black P under pressure.

The article is organized as follows. In Section II, as also
supplemented by the appendices, we introduce the detailed formalism of
our {\it ab initio} analysis for the electronic and crystal structure
as well as the {\eph} coupling. For the former, as opposed to previous
studies, we employ full potential calculations. For the latter, we
likewise refine previous studies by considering band-selective {\eph} coupling strengths. Section III proceeds
by giving a detailed account on the results of our analysis for simple
cubic black P. This also includes a detailed orbital-resolved study of the
electronic structure, as well as the calculation of {\eph} coupling
via different {\it ab initio} approaches. As we are bringing together
our theoretical findings with the experimental evidence in Section IV,
we can quantitatively and qualitatively rationalize the {\Tc} valley
formation in the simple cubic phase. In Section V, we conclude that
while we are still short of an answer to some of the experimental
features, in particular to those observed for higher pressures, our
systematic study supports our resulting hypothesis that electronic {\it ab
  initio} approaches are insufficient to cover the full phenomenology
of black P. Instead, we suggest that electronic correlations may enter
within a degree of sophistication that is beyond such methods, and
that plasmonic contributions are likely to improve the current
theoretical understanding of the material.

\section{Methods}

\subsection{Electron phonon coupling formalism}

We start by describing the {\eph} coupling formalism we employ. The
{\eph} matrix element for the scattering of an electron in band $n$ at
wave vector $\textbf{k}$ to a state in band $m$ with wave vector
$\textbf{k}+\textbf{q}$, with the absorption or emission of a phonon
with mode $\nu$ at wave vector $\textbf{q}$, is
\begin{equation}
\label{g}
g^{\nu}_{mn}(\textbf{k},\textbf{q})=\sqrt{ \frac{\hbar}{2M\omega_{{\bf q}\nu}}}\langle m,\textbf{k}+\textbf{q}|\delta_{\textbf{q}\nu}V_{SCF}|n,\textbf{k}\rangle\,.
\end{equation}
In \eqref{g}, $|n,\textbf{k}\rangle$ is the electronic Bloch
state, $\omega_{{\bf q}\nu}$ is the screened phonon frequency, $M$ is
the atomic mass, and $\delta_{\textbf{q}\nu}V_{SCF}$ is the derivative
of the self-consistent potential with respect to the collective atomic
displacements corresponding to phonon mode $\nu$ at wave vector
${\bf q}$. The phonon self energy is given by
\begin{equation}
\Pi_{{\bf q}\nu}=\frac{1}{N_{\bm{k}}}\sum_{mn,{\bf k}}|g^{\nu}_{mn}(\textbf{k},\textbf{q})|^2\frac{n_{\rm F}(\epsilon_{m{\bf k}+{\bf q}})-n_{\rm F}(\epsilon_{n{\bf k}})}{\omega_{{\bf q}\nu}+i\delta+\epsilon_{m{\bf k}+{\bf q}}-\epsilon_{n{\bf k}}}\,,
\end{equation}
where $\epsilon_{n{\bf k}}$ is the energy relative to the Fermi level
for the Bloch state, $n_{\rm F}(\epsilon)$ is the Fermi distribution
function and $N_{\bm{k}}$ is the number of $k$ points. The phonon
linewidth is proportional to the imaginary part of the phonon self
energy. Considering the large value of the Fermi energy compared to
the phonon frequencies, the phonon linewidth at low temperature can be
written as\cite{Allen1972}
\begin{equation}
\gamma_{{\bf q}\nu}(\omega)=\frac{2\pi\omega_{{\bf q}\nu}}{N_{\bm{k}}}\sum_{mn,{\bf k}}|g^{\nu}_{mn}(\textbf{k},\textbf{q})|^2\delta(\epsilon_{m{\bf k}+{\bf q}})\delta(\epsilon_{n{\bf k}})\,.
\label{phline}
\end{equation}
The {\eph} coupling constant for a specific phonon mode is,
\begin{equation}
\lambda_{{\bf q}\nu}=\frac{\gamma_{{\bf q}\nu}}{\pi N(E_{\rm F})\omega^2_{{\bf q}\nu}}\,,
\label{lambda1}
\end{equation}
where $N(E_{\rm F})$ is the density of states per spin at the Fermi
level. In terms of the phonon linewidths, the Eliashberg spectral
function $\alpha^2F(\omega)$ can be written as
\begin{equation}
\alpha^2F(\omega)=\frac{1}{2\pi N(E_{\rm F})}\sum_{{\bf q}\nu}\frac{\gamma_{{\bf q}\nu}}{\omega_{{\bf q}\nu}}\delta(\omega-\omega_{{\bf q}\nu})\,.
\end{equation}
Finally, the isotropic {\eph} coupling constant is defined as
\begin{equation}
\lambda=2\int^{\infty}_{0} d\omega\, \frac{\alpha^2F(\omega)}{\omega}.
\label{IntegratedL}
\end{equation}
For multiband systems, as is the case for simple cubic P, it is useful to
introduce a band-resolved {\eph} coupling constant $c_{mn}$, which
describes the Cooper pair scattering of an electron from band $m$ to
band $n$ by phonons, and is defined as\cite{Liu2001}
\begin{equation}\begin{split}
c_{mn}&=V_{mn}N_mN_n \\
&=\frac{2}{N_{\bm{q}}N_{\bm{k}}} \!\!\sum_{\bm{k}\in n, \bm{k}+\bm{q}\in m,\nu}\omega^{-1}_{{\bf q}\nu}|g^{\nu}_{mn}(\textbf{k},\textbf{q})|^2\delta(\epsilon_{m{\bf k}+{\bf q}})\delta(\epsilon_{n{\bf k}})\,,
\label{phline1}
\end{split}\end{equation}
where $N_{m}$ ($N_{n}$) is the DOS at the Fermi level contributed by
$m$-th ($n$-th) band and $V_{mn}$ is the effective interaction for
Cooper pair scattering. The isotropic {\eph} coupling
constant can be written as
\begin{equation}
\lambda=\sum_{mn}\frac{c_{mn}}{N(E_F)}=\sum_{mn}\frac{V_{mn}N_mN_n}{N(E_F)}\,.
\end{equation}
In multiband systems, the {\eph} constant is given by the maximum
eigenvalue $\lambda_{\rm multi}$ of the matrix $\Lambda$, which is defined
as
\begin{equation}
\Lambda_{mn}=V_{mn}N_n=\frac{c_{mn}}{N_m}\,.
\end{equation}
The band-resolved Eliashberg function $\alpha^2F_{mn}(\omega)$ reads
\begin{equation}\begin{split}
\alpha^2&F_{mn}(\omega) \\
&=\!\!\!\sum_{\bm{ k}\in n, \bm{k}+\bm{q}\in m,\nu} \!\!\!\!\!\!\!\!\!\!\frac{|g^{\nu}_{mn}(\textbf{k},\textbf{q})|^2\delta(\epsilon_{m{\bf
    k}+{\bf q}})\delta(\epsilon_{n{\bf k}})\delta(\omega-\omega_{{\bf
    q}\nu})}{N_n N_{\bm{q}}N_{\bm{k}}}\,. \label{eq:alpha2f}
\end{split}\end{equation}
Then, the matrix element $\Lambda_{mn}$ can be written as
\begin{equation}
\Lambda_{mn}=2\int^{\infty}_{0}d\omega\frac{\alpha^2F_{mn}(\omega)}{\omega}\,.
\end{equation}
{\Tc} can be determined using the Allen-Dynes-modified McMillan
equation\cite{McMillan1968,Allen1975},
\begin{equation}
T_{\rm c} =\frac{\omega_{ln}}{1.2}\exp\bigg\{-\frac{1.04(1+\lambda)}{\lambda-\mu^*(1+0.62\lambda)}\bigg\}\,,
\label{mc}
\end{equation}
where $\mu^*$ is the Coulomb pseudopotential\cite{Morel1962} and
$\lambda$ is the isotropic (or multi-band) {\eph} coupling
constant. $\omega_{ln}$ is defined as
\begin{equation}
\omega_{ln}=\exp\bigg\{\frac{2}{\lambda}\int^{\infty}_{0}d\omega\,\ln\omega\frac{\alpha^2F(\omega)}{\omega}\bigg\}\,.
\end{equation}
Using Eq.~\ref{mc}, we obtain {\Tc} based on a suitable choice of $\mu^*$ and
compare with previous works. However, we also use the solution of the
isotropic Eliashberg equations to determine {\Tc} without any empirical
parameters\cite{Allen1960,Sanna2012}.

\subsection{Details of the calculations}

In the relaxation with external pressure, we employ the projector
augmented wave (PAW) method~\cite{Bloechl1994} as implemented in the
Vienna ab initio simulation package
(VASP)\cite{Kresse1993,Kresse1996,Kresse1996prb}, where both the local
density approximation (LDA) and the generalized-gradient approximation
(GGA)\cite{Perdew1996} for the exchange correlation functional are
used. The expansion of the wave functions in the plane-wave basis is
cut off at 600 eV. The adopted $k$-point mesh is $37\times37\times37$
and the convergence criterion is that all the forces on atoms are less
than 0.001~eV/\AA. The {\eph} coupling calculations are
performed with the full-potential linear augmented plane wave
(FP-LAPW) method\cite{Andersen1975} as implemented in the Elk
code\cite{elk,Dewhurst2003}. In the calculations, we chose the
Perdew-Burke-Ernzerhof (PBE) exchange correlation functional. The
adopted muffin-tin radii $R_{\rm MT}$ (2.05-2.20 a.u.) for P depend on
the lattice constants and the plane-wave cutoff is set to $R_{\rm
  MT}\times k_{\rm max}=8.5$. The angular momentum expansion in the
muffin tins is taken to $l_{\rm max}=12$ for both the wave functions
and the potential. We use the full potential local orbital (FPLO)
basis~\cite{Koepernik1999} as a further accuracy check for the
electronic structure. For accurate DOS calculations, we use
$\bm{k}$-point meshes of $100\times100\times100$. Phonon dispersions
are obtained by using the supercell method, and we have carefully
checked the convergence of the phonon spectrum with respect to $q$-point
mesh. We find that an $8\times8\times8$ $q$-point mesh is enough to
obtain the converged phonon spectrum. For the phonon linewidth in
Eq.~\ref{phline}, in Elk the $\delta$ function is numerically replaced
by a Gaussian function with a smearing parameter $\sigma$. The
convergence of $\gamma_{\nu\bm{q}}$ with respect to $k$-point mesh has
also been carefully checked, and the final results are obtained from
calculations with a $128\times128\times128$ $k$-point mesh and
$\sigma=0.004\,{\rm Ha} \approx 100\,{\rm meV}$. With the chosen
parameters, $\lambda$ is estimated to converge within 0.01 (i.e. less
than 2\%). We also perform {\eph} calculations to obtain the
resistivity by using the EPW (short for electron-phonon coupling using
Wannier functions) method\cite{Giustino2007,Ponce2016}, as implemented
in the Quantum Espresso pseudopotential plane wave code
package\cite{Giannozzi2009}.

\section{Results}

\subsection{Crystal and electronic structure for simple cubic
  phosphorous}

\begin{figure}[tb]
\includegraphics[width=\columnwidth]{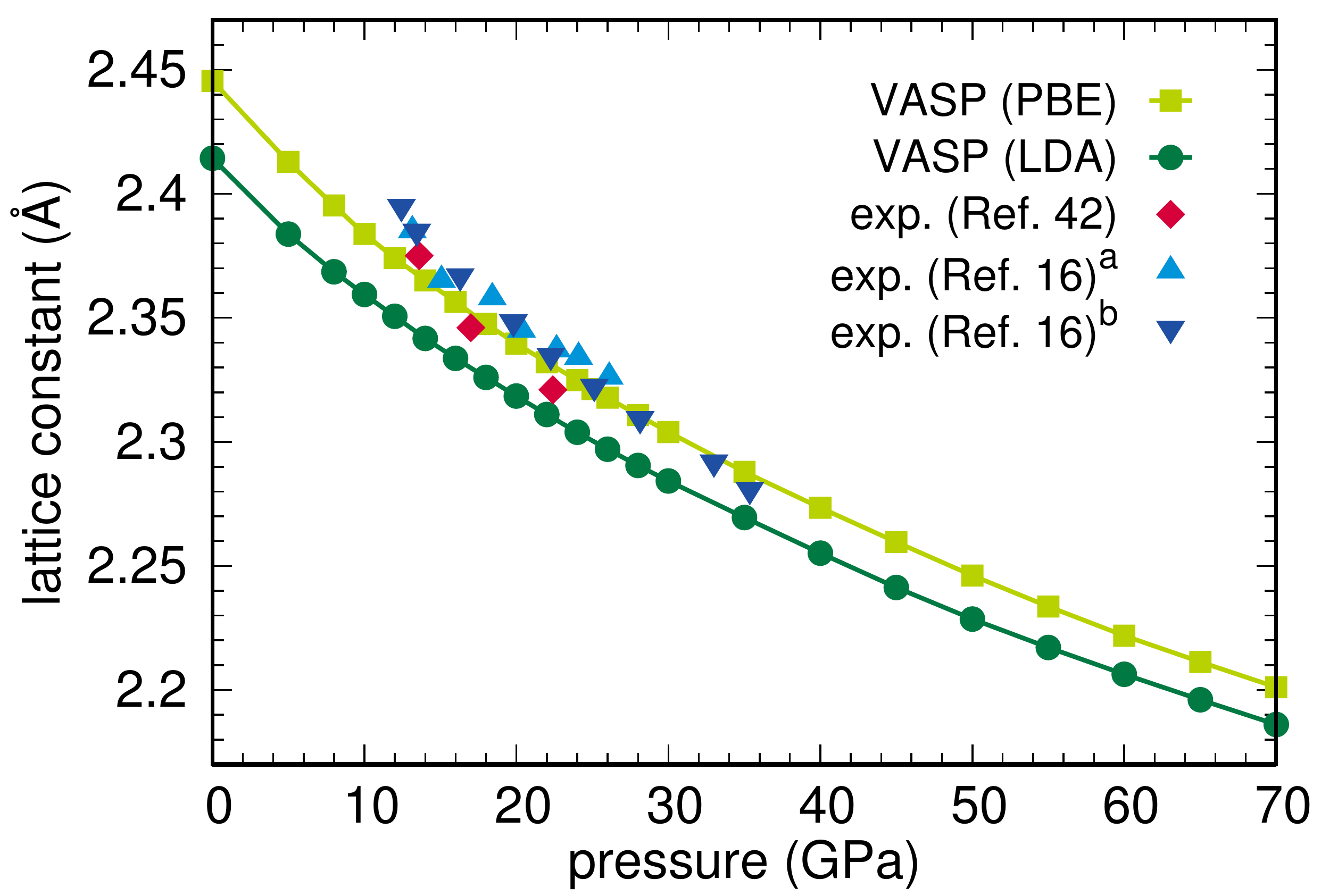}
\caption{(Color online) Theoretical and experimental lattice constants
  as a function of external pressure for simple cubic
  phosphorous. Calculations are performed with VASP using two
  different exchange correlation functionals, and experimental data
  are taken from Ref.~\onlinecite{Clark2010} and
  Ref.~\onlinecite{Guo2016} ($^{\rm a}$ and $^{\rm b}$ refer to two
  different runs).  \label{ados_pressure}}
\end{figure}

\begin{figure}[tb]
\includegraphics[width=\columnwidth]{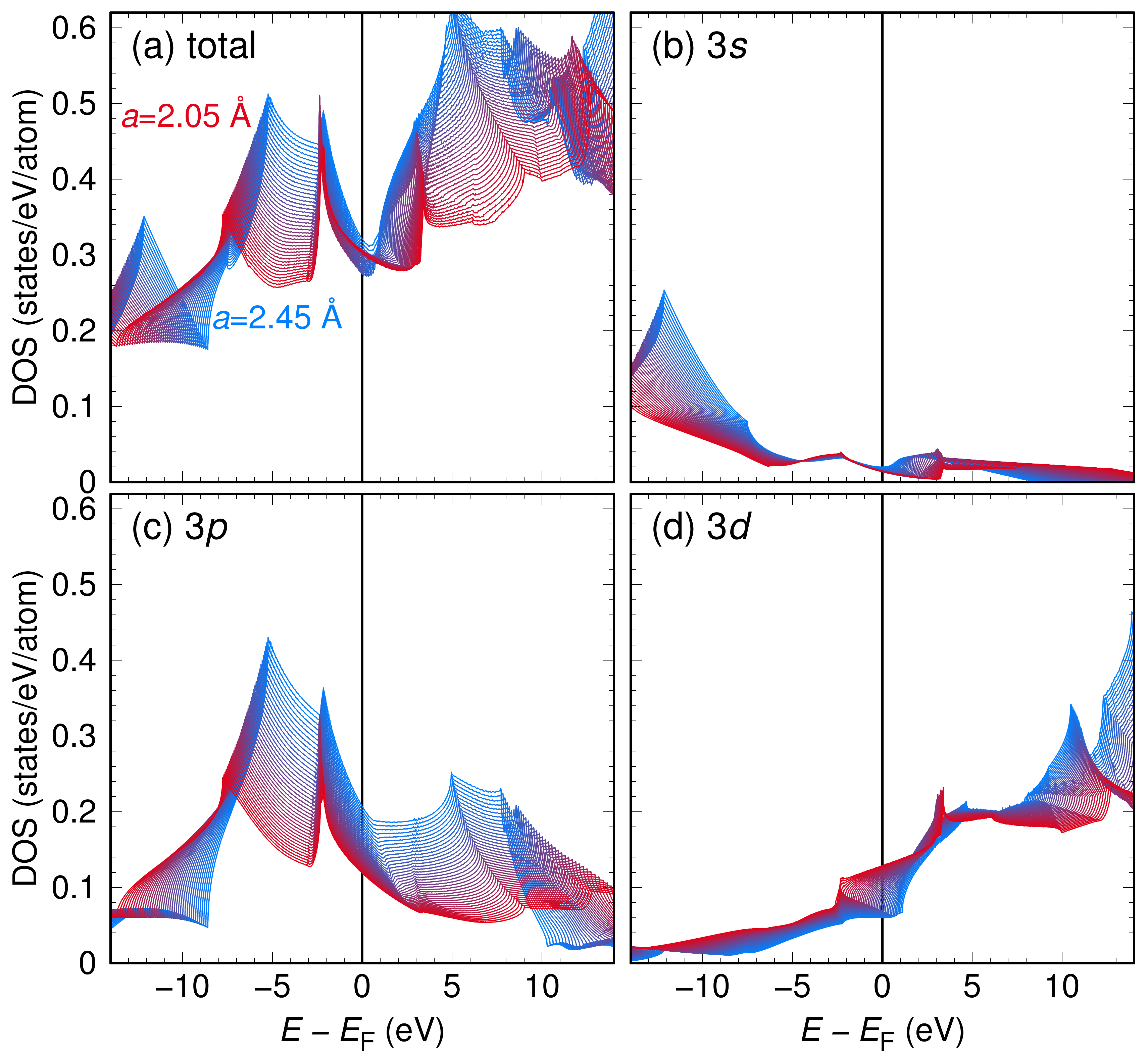}
\caption{(Color online) Total and orbital projected densities of
  states in the simple cubic phase for various lattice constants. This
  calculation was done with FPLO. The gradient lines ranging from blue to red represent the increasing of pressures (decreasing of lattice constantsp). \label{dos_pressure}}
\end{figure}

\begin{figure}[tb]
\includegraphics[width=\columnwidth]{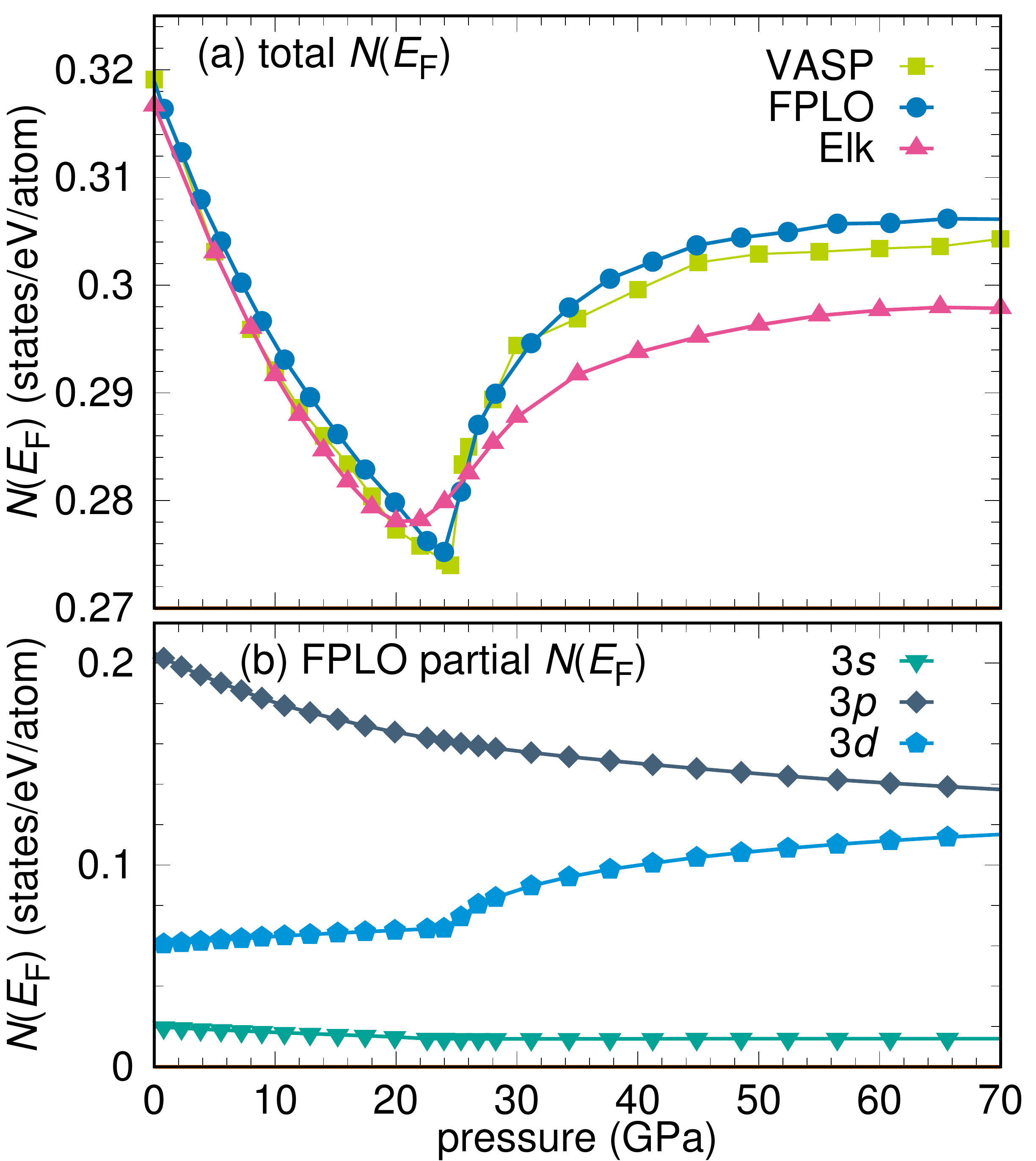}
\caption{(Color online) Density of states at the Fermi level for the
  simple cubic phase as a function of external pressure. (a)
  Comparison between different basis sets. (b) Orbital contributions
  to $N(E_{\rm F})$. \label{Efdos_pressure}}
\end{figure}

\begin{figure}[tb]
\includegraphics[width=\columnwidth]{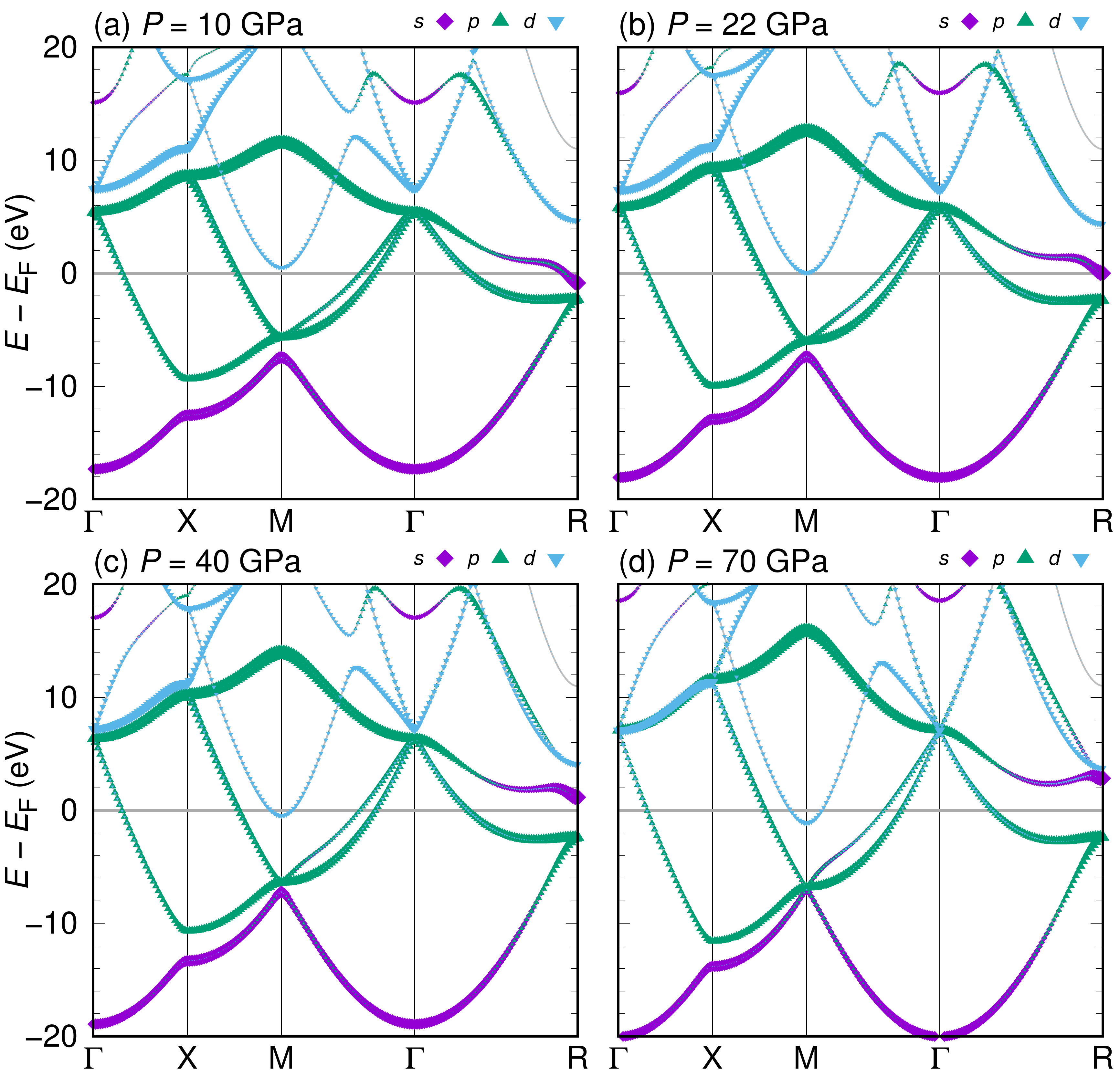}
\caption{(Color online) Band structures with orbital weights for
  simple cubic phosphorous at four different external
  pressures.  \label{bandstructure}}
\end{figure}

\begin{figure}[tb]
\includegraphics[width=\columnwidth]{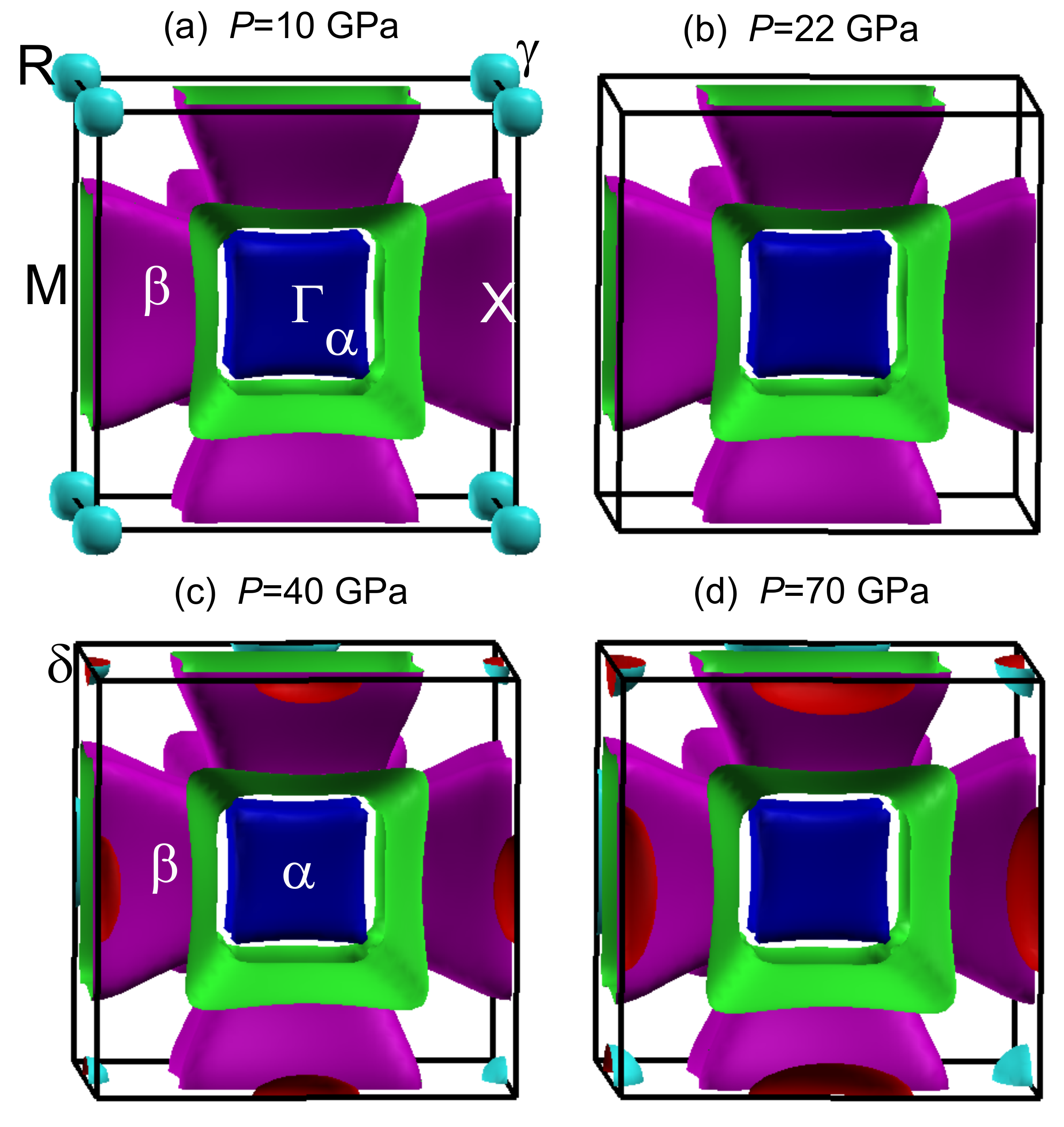}
\caption{(Color online) The Fermi surfaces for external pressures
  $P$=10 (a), 22 (b), 40 (c) and 70 GPa (d).  \label{FS}}
\end{figure}

The calculated lattice constants as a function of external pressure
for simple cubic phosphorus are shown in Fig.~\ref{ados_pressure}, in
comparison with experimental data\cite{Clark2010,Guo2016}. The lattice
constants from PBE are always slightly larger than those of LDA. From
the figure, we find that the calculated lattice constants from PBE
matches better with available experimental data, suggesting that the PBE
functional may be more suitable for P than the LDA
functional. Therefore, we adopted the PBE functional in all further
calculations. The total density of states (DOS) and the orbital
projected DOS are shown in Fig.~\ref{dos_pressure}. At ambient
pressure, the states near the Fermi level $E_{\rm F}$ are mainly
formed by $p$-orbitals. As the pressure increases, however, the total
states display shifts to lower energy, and the contribution from
$p$-orbital and $s$-orbital states at $E_{\rm F}$ decreases while the
$d$-orbital weight increases. The DOS at the Fermi level $N(E_{\rm
  F})$ as a function of external pressure is shown in
Fig.~\ref{Efdos_pressure}~(a). We find that $N(E_{\rm F})$ first
decreases linearly, and then increases with the increase of external
pressure, which results in a local minimum in $N(E_{\rm F})$. The
corresponding pressure is 22~GPa ($a=2.332$~{\AA}) in PBE
calculations. The $N(E_{\rm F})$ we obtain as function of pressure in
principle compares well with previous
calculations\cite{Rajagopalan1989,Cohen2013} but has a more pronounced
minimum. The orbitally resolved $N(E_{\rm F})$ in
Fig.~\ref{Efdos_pressure}~(b) shows that the $3d$ orbital is
responsible for the increase of $N(E_{\rm F})$ beyond the minimum at
$P=22$~GPa; taking only $s$ and $p$ orbitals into account, $N(E_{\rm
  F})$ would fall monotonously.

To explain the observed changes in the DOS, we plot the orbital
projected band structures for external pressures of $P=10$~GPa,
22~GPa, 40~GPa and 70~GPa in Fig.~\ref{bandstructure}. We find that
the major changes at the Fermi level occur in bands with $s$ and $d$,
rather than $p$ character. At the $R$ point, a band with $s$ character
which is partially occupied below 22~GPa shifts up in energy with
increasing pressure, while a band around $M$ with $d$ character
changes with pressure in an opposite fashion, and starts to become
populated around $P=22$~GPa. These changes are consistent with
previous calculations\cite{Aoki1987,Rajagopalan1989,Cohen2013}. The
corresponding Fermi surfaces (FSs) at these pressures are shown in
Fig.~\ref{FS}. At 10 GPa, close to the rhombohedral-simple cubic phase
transition point in experiment, there are three FSs: a cubic hole FS
$\alpha$ around $\Gamma$, a big open-type FS $\beta$ and a small
spherical electron FS $\gamma$. The $\alpha$ and $\beta$ FSs are
attributed to $p$ orbitals and enclose about 0.07 holes and 1.068
electrons (or 0.932 holes), respectively. The band contributing to the
$\beta$ FS is close to half filling. The $\gamma$ FS, predominantly
formed by $s$ orbitals, encloses about 0.002 electrons. As pressure
increases, the $\gamma$ FS gradually shrinks and the $\beta$ FS grows
a bit while the $\alpha$ FS remains almost unchanged. At $P=22$~GPa,
the $\gamma$ FSs disappear completely, and the Lifshitz transition
occurs. With a slight further increase of pressure, three ellipsoidal
electron FSs $\delta$ appear around $M$. Two Lifshitz transitions
happen almost at the same pressure and, as it turns out, have a
significant effect on the {\eph} coupling. The $\delta$ FSs are of $d$
orbital character, as expected from the bandstructures in
Fig.~\ref{bandstructure}. As correlation effects in $d$ orbitals are
stronger as compared to $p$ orbitals, the appearance of the $\delta$
FSs around $M$ may trigger stronger electronic correlation
effects. This runs contrary to the expectation that pressure increases
the bandwidth, and thus should decrease correlation effects as a
matter of principle. To investigate the correlation effects, we
performed calculations with hybrid functional and GGA+U. The band
structures are given in Appendix \ref{RCphase}. In calculations with
the hybrid functional, the bandwidth increases and the most noticeable
changes near the Fermi level are that the $s$ band at $R$ and the $d$
band at $M$ are are shifted to slightly higher energy, and as a
consequence the two Lifshitz transitions happen separately. In GGA+U
calculations, however, the obtained band structures are almost
unchanged compared to the GGA functional. Therefore, correlations have
little effect on the band structure, but may still have a significant
effect on the superconductivity of phosphorus.

\subsection{Electron phonon coupling for cubic phosphorous}

\begin{figure}[tb]
\includegraphics[width=\columnwidth]{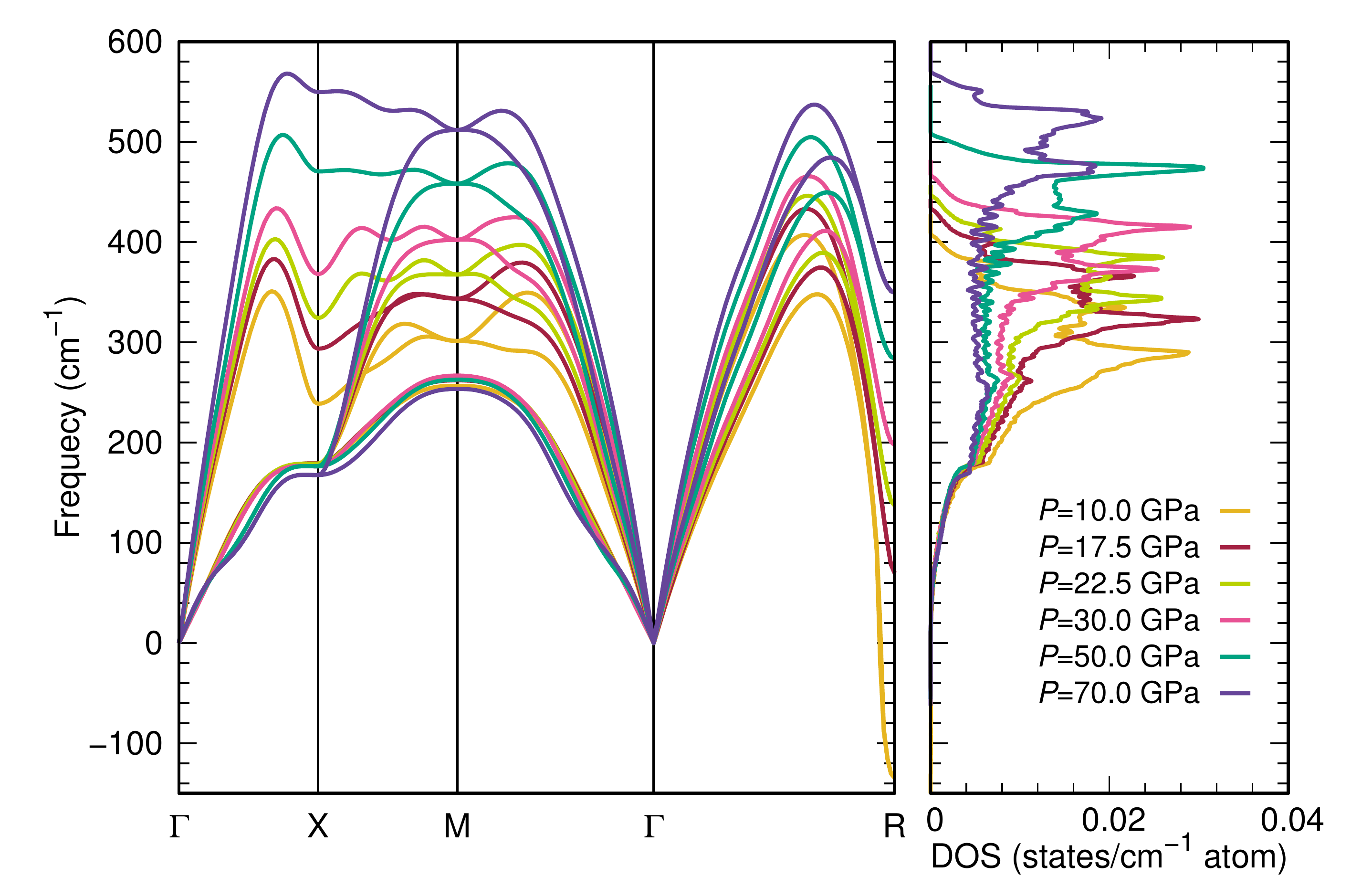}
\caption{(Color online) Phonon spectra and densities of states for
  simple cubic phosphorus at several pressures.  \label{phdisdos}}
\end{figure}
Now we turn to the effect of pressure on the phonons for simple cubic
phosphorus. The phonon spectra and densities of states are shown in
Fig.~\ref{phdisdos} for several pressures. Below $P=16$~GPa, there are
imaginary phonon modes around the $R$ point. The unstable modes
correspond to the lattice distortion in the direction of the
rhombohedral $A7$ phase\cite{Cohen2013}. Above $P=16$~GPa, unstable
phonon modes vanish and most phonon states lie in two regions: A
low-energy region (LER) and a high-energy region (HER). DOS is small
in the LER but shows large peaks in the HER; the latter contains more
states than the former. At 17.5 GPa, for example, the LER is from 180
cm$^{-1}$ to 280 cm$^{-1}$ and the HER is from 280 cm$^{-1}$ to 400
cm$^{-1}$. In the LER, phonon bands are relatively smooth, and phonon
states are mainly due to phonon modes around $\Gamma$ and R as well as
low-energy modes around $X$ and $M$. In the HER, the situation is quite
different. Along the $\Gamma-X$ direction, the frequency of the
longitudinal mode increases and reaches a maximum near the $X$ point, to
then drop suddenly. One mode along $\Gamma-M$ and three modes along
$\Gamma-R$ direction exhibit similar behavior, and the drops near the
$R$ point are the largest. All this finally results in a large DOS in the
HER. As pressure increases, the drops near $X$, $M$, and $R$ decrease
and most phonon modes harden, which leads to the widening of the LER
and a shift towards higher frequency for the HER. These changes have a
significant effect on the {\eph} coupling discussed in the
following.

\begin{figure}[tb]
\includegraphics[width=\columnwidth]{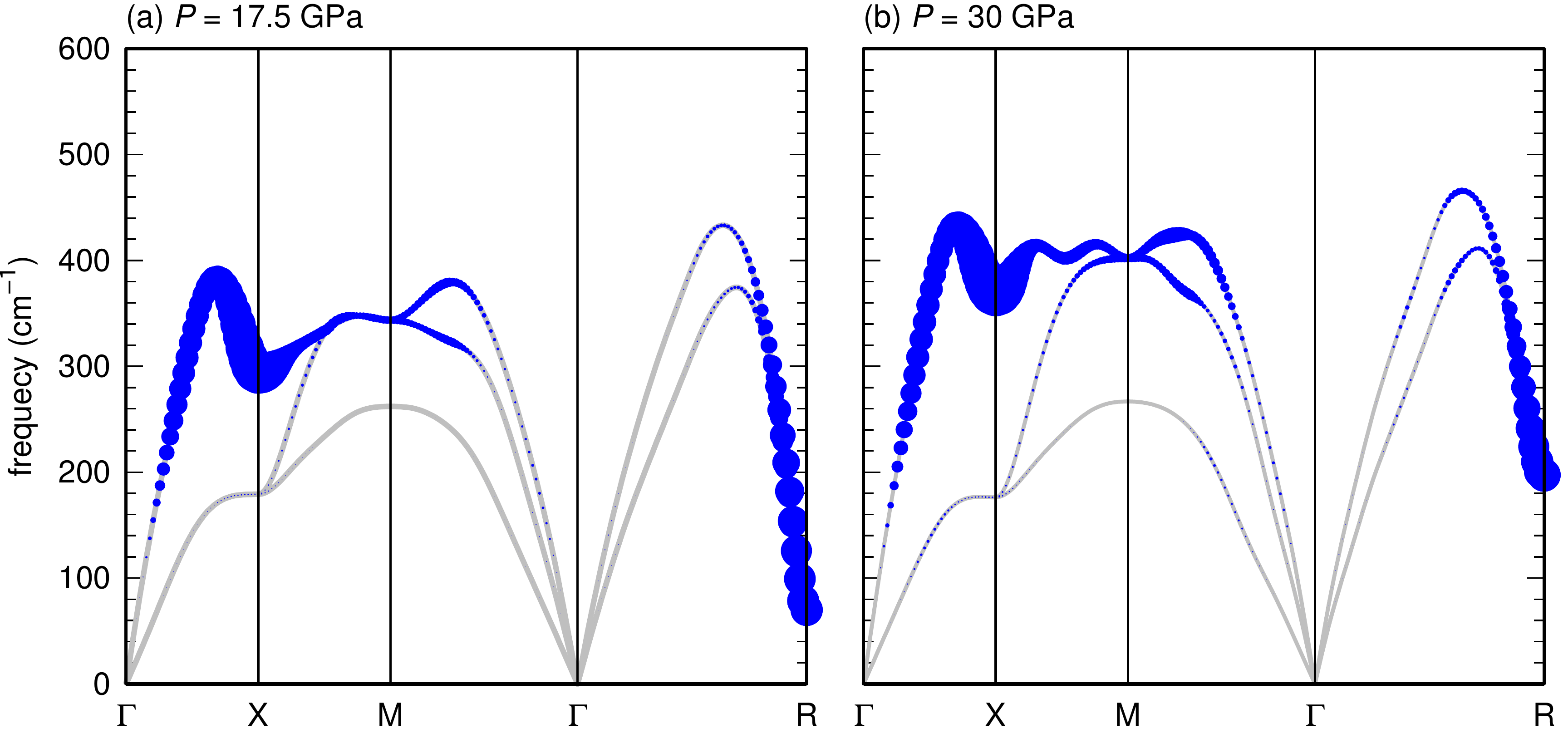}
\caption{(Color online) The phonon linewidth $\gamma_{\nu\bm{q}}$
  along high symmetry $q$-points at P=17.5 GPa (before the Lifshitz
  transition) and P=30 GPa (after the Lifshitz transition). The sizes
  of the circles represent the magnitude of
  $\gamma_{\nu\bm{q}}$.  \label{gamma}}
\end{figure}

\begin{figure}[tb]
\includegraphics[width=\columnwidth]{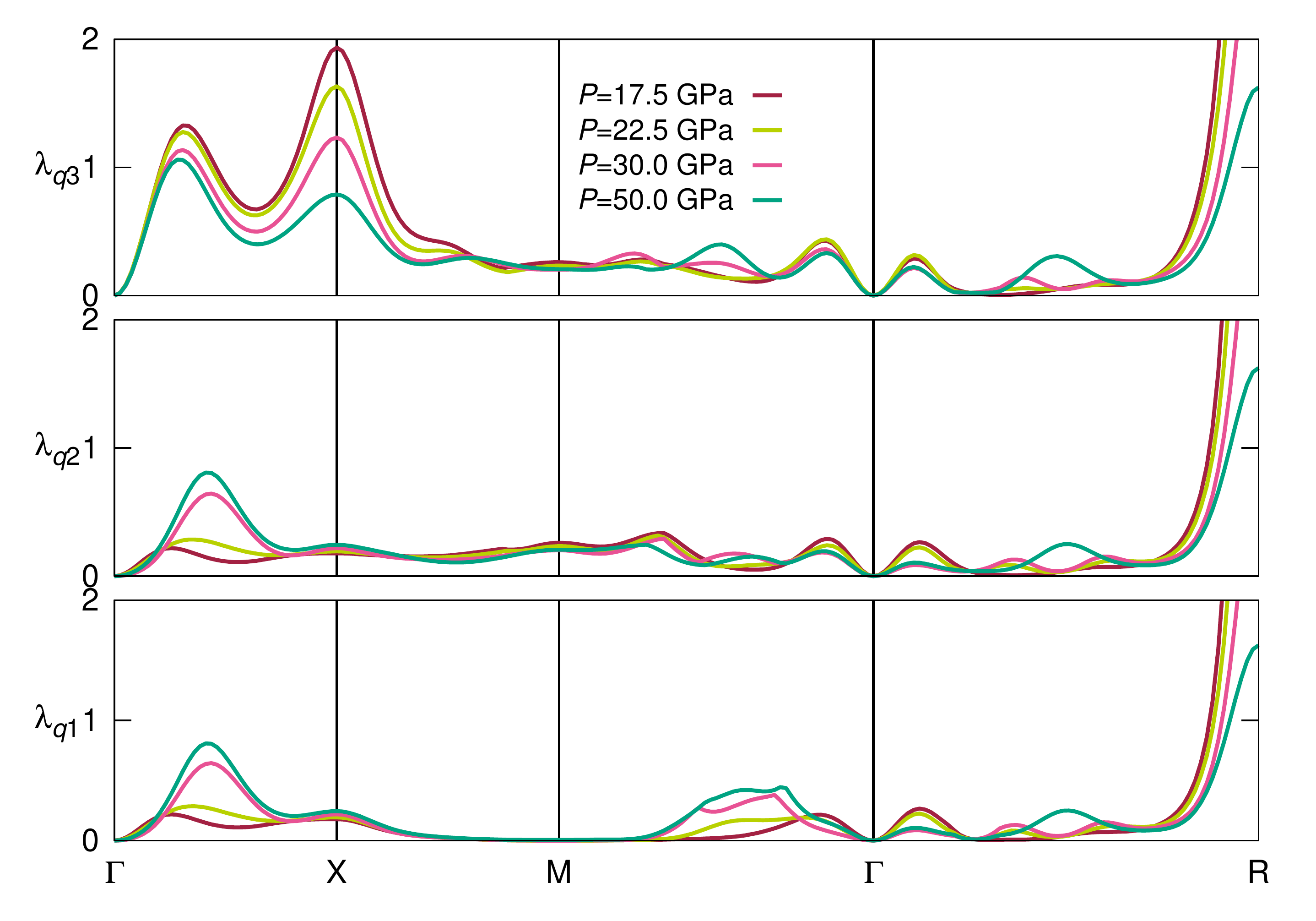}
\caption{(Color online) Electron phonon coupling constant
  $\lambda_{\nu\bm{q}}$ along high symmetry $q$ directions at
  $P=17.5$~GPa, 22.5~GPa, 30.0~GPa and 50.0~GPa.  \label{lambda}}
\end{figure}

To investigate the {\eph} coupling for simple cubic phosphorus, we
first plot the phonon linewidth at two instructive values of pressure
(before and after the Lifshitz transitions) in
Fig.~\ref{gamma}. Strong {\eph} coupling occurs in the high-energy
modes around $X$ and $M$ and all modes around $R$, which is due
to the large Fermi-surface nesting at these $\bm{q}$
vectors\cite{Cohen2013}. With the increase of pressure, the phonon
linewidth near $X$, $M$ and $R$ increases a little bit.  The obtained
$\lambda_{\nu\bm{q}}$ along high symmetry $k$-points for four
pressures are shown in Fig.~\ref{lambda}. The {\eph} coupling constant
$\lambda_{\nu\bm{q}}$ is peaked at R. It is rather different from the
distribution of phonon linewidths $\gamma_{\nu\bm{q}}$, where
$\gamma_{3M}$ and $\gamma_{3X}$ are comparable with $\gamma_{R}$. As
the phonon frequencies at $R$ are lower than those at $X$ and
$M$, and since $\lambda_{\nu\bm{q}}$ is inversely proportional to
$\omega_{\nu\bm{q}}^2$, the $\lambda_X$ and $\lambda_M$ are much
smaller than $\lambda_R$. Most $\lambda_{\nu\bm{q}}$ decrease with
increasing pressure. However, contrary to the pressure-dependent
behavior of $\lambda_{\nu\bm{q}}$ around $X$, $M$ and $R$, after the
Lifshitz transitions the $\lambda_{1\bm{q}}$ and $\lambda_{2\bm{q}}$
around $\Gamma$ (along $\Gamma-X$, $\Gamma-M$ and $\Gamma-R$
directions) increase with increasing pressure, which can be attributed to
the softening of phonon modes at these $\bm{q}$ vectors. These
features are consistent with the previous plane wave pseudopotential
calculation\cite{Cohen2013}.

\begin{figure}[tb]
\includegraphics[width=\columnwidth]{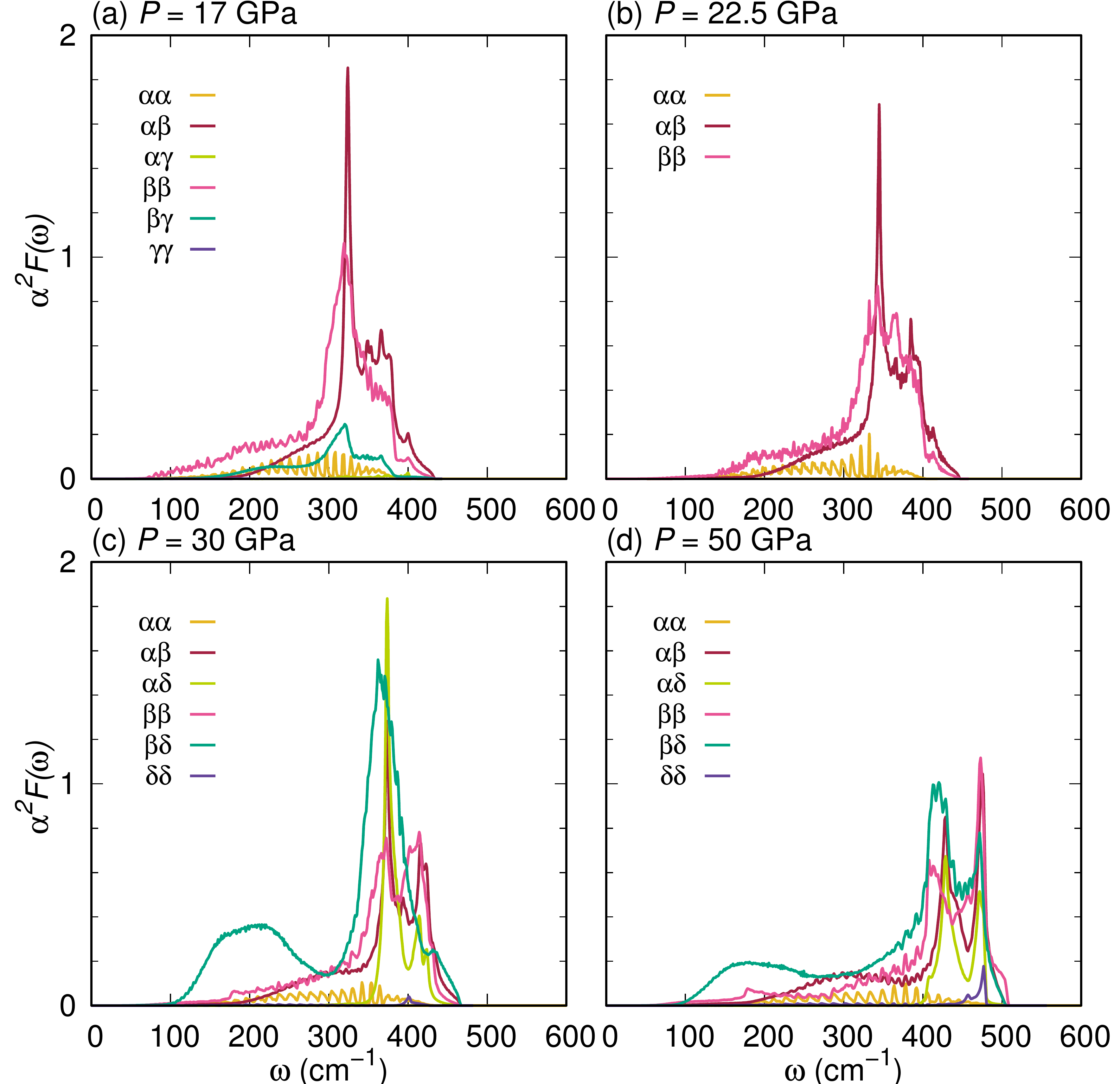}
\caption{(Color online) Band-resolved Eliashberg spectral function
  $\alpha^2F_{mn}(\omega)$ for cubic phosphorus at
  $P=17.5$~GPa, 22.5~GPa, 30.0~GPa and 50.0~GPa. \label{bandalpha2f}}
\end{figure}

\begin{figure}[tb]
\includegraphics[width=\columnwidth]{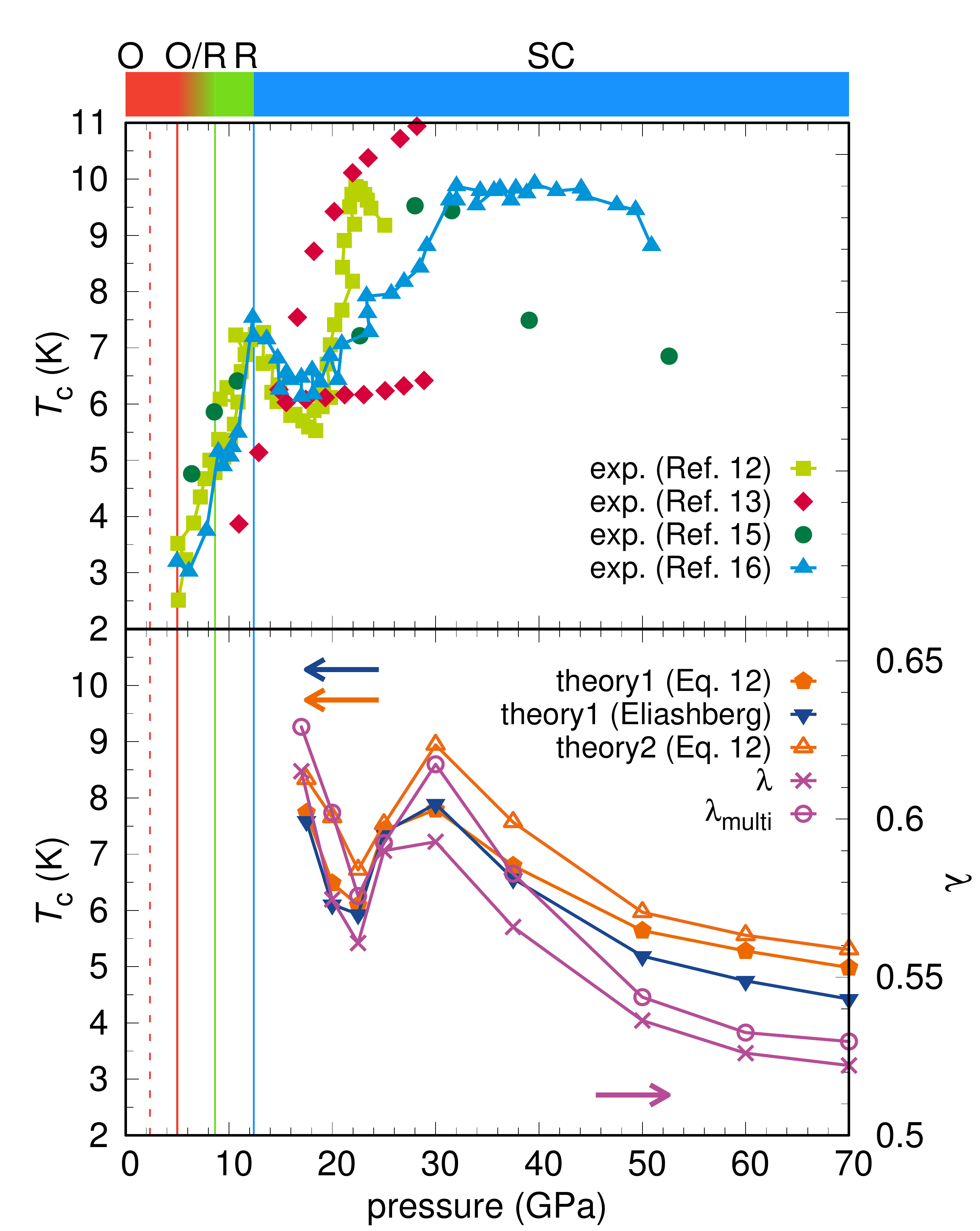}
\caption{(Color online) Superconducting transition temperature {\Tc}
  and {\eph} coupling constant $\lambda$ as a function of
  pressure in theoretical calculations compared to the experimental
  data. Squares, circles, diamonds and triangles denote the data from
  J. Wittig \emph{et al.}\cite{Wittig1985}, Karuzawa \emph{et
    al.}\cite{Karuzawa2002}, Kawamura \emph{et al.}\cite{Kawamura1984}
  and Guo \emph{et al.}\cite{Guo2016}, respectively. "theory1" denotes
  {\Tc} from the isotropic $\lambda$ and "theory2" denotes {\Tc} from
  the multi-band $\lambda_{\rm multi}$. \label{lambdatc} } 
\end{figure}

In order to analyze the contributions of different Fermi surfaces to
the {\eph} coupling, we plot the band resolved Eliashberg spectral
functions in Fig.~\ref{bandalpha2f}. Before the Lifshitz transition, at
$P=17.5$ GPa the $\beta\beta$ intraband coupling and $\alpha\beta$
interband coupling contribute dominantly to {\eph} coupling, which is
due to the large 
$\beta$ Fermi surfaces and relative good nesting between $\alpha$ and
$\beta$ Fermi surfaces. As the spherical $\gamma$ FS is well separated
from $\alpha$ and $\beta$ FS in $k$-space, the $\alpha\gamma$ and
$\beta\gamma$ interband couplings are relatively weak. Upon increasing
pressure, the peak of $\beta\beta$ intraband coupling becomes broad
and shifts to higher frequency, which results in a significant {\eph}
coupling decrease in the $\beta\beta$ channel. At the Lifshitz
transition point, the $\alpha\gamma$ and $\beta\gamma$ interband
couplings vanish due to the disappearance of the $\gamma$ FS. Thus, {\eph}
coupling decreases with increasing pressure before the Lifshitz
transition, as shown in Fig.~\ref{lambdatc}. As pressure further
increases, three $\delta$ FSs with $d$ orbital character appear around
the $M$ point, and they introduce strong interband couplings in
$\alpha\delta$ and $\beta\delta$ channels (see
Fig.~\ref{bandalpha2f}~(c)), similar to the case of sulfur under
pressure\cite{Monni2017}. The {\eph} coupling increases with
increasing pressure because the {\eph} coupling enhancements from new
channels overcome the weakening of couplings in $\alpha\beta$ and
$\beta\beta$ channels. Therefore, $\lambda$ exhibits a valley around the
Lifshitz transition point. After the Lifshitz point, the
isotropic $\lambda$ increases and reaches a maximum at $P=30$ GPa. As
pressure further increases, the two broad peaks in $\beta\delta$
channels show opposite behaviors: one shifts to higher frequency and
the other shifts to lower frequency. The remaining interband and
intraband {\eph} couplings shift to higher frequency. This will lead
to decreased {\eph} couplings in most channels except the
$\delta\delta$ intraband channel. Therefore, $\lambda$ will decrease
again.

\begin{figure}[tb]
\includegraphics[width=\columnwidth]{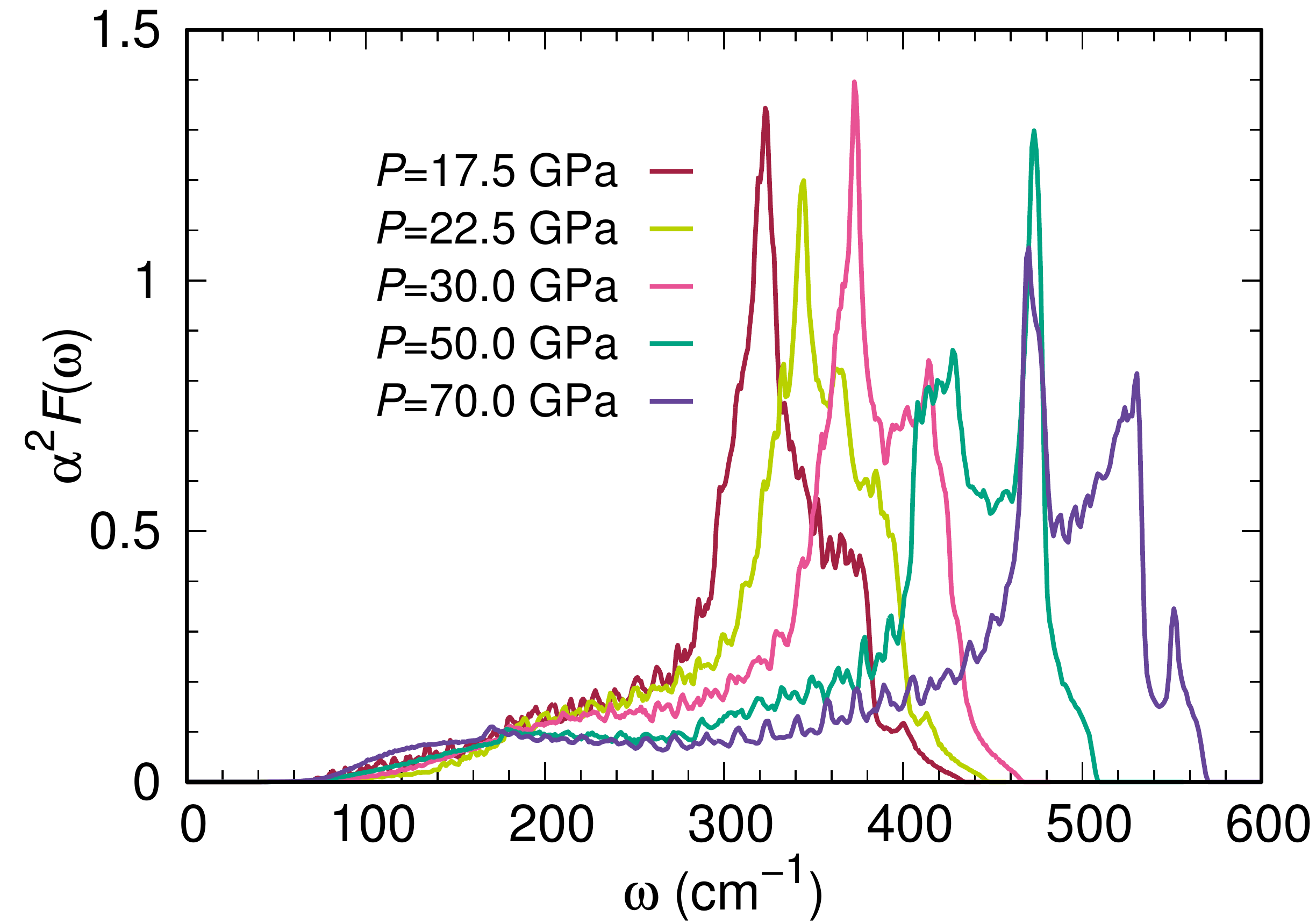}
\caption{(Color online) Eliashberg spectral function
  $\alpha^2F(\omega)$ for cubic phosphorus at $P=17.5$~GPa ( $a$=2.350
  \AA~), 22.5~GPa ( $a$=2.330 \AA~), 30.0~GPa ( $a$=2.300 \AA~),
  50.0~GPa ( $a$=2.246 \AA~) and 70.0~GPa ( $a$=2.201
  \AA~).  \label{alpha2f}}
\end{figure}

The isotropic Eliashberg spectral functions $\alpha^2F(\omega)$ at
five selected pressures are shown in Fig.~\ref{alpha2f}. We find that the
$\alpha^2F(\omega)$ have almost the same shapes as the phonon DOS
shown in Fig.~\ref{phdisdos}. The {\eph} coupling occurs mainly in the
high energy region and it is mainly due to $\beta\beta$ intraband
coupling and $\alpha\beta$, $\alpha\delta$ and $\beta\delta$ interband
coupling. This high energy spectral weight shifts to higher frequency
with increasing pressure, rather similar to the behavior of the phonon
DOS. Another noticeable feature in $\alpha^2F(\omega)$ is that the
{\eph} coupling (100$cm^{-1}$ to 200$cm^{-1}$) in the low energy
region increases as pressure increases, which can be mainly attributed
to {\eph} coupling in the $\beta\delta$ channel.

We calculate {\Tc} using the McMillan equation (Eq.~\ref{mc}) and the
Coulomb pseudopotential parameter $\mu^{*}$ is fixed to $\mu^*$=0.12, a
a value for $\mu^{*}$ that matches the minimum {\Tc} in our calculations to the experimentally determined {\Tc} valley observed in Ref. \onlinecite{Guo2016}.
This $\mu^*$ is significantly smaller than that in Ref.~\onlinecite{Cohen2013}. We
also calculate {\Tc} by solving the isotropic Eliashberg equations
self-consistently\cite{Allen1960}. Fig.~\ref{lambdatc} shows the
calculated isotropic {\eph} coupling constant $\lambda$, multi-band
{\eph} coupling constant $\lambda_{\rm multi}$, and {\Tc} as a
function of pressure in theory, compared to the experimental data from
different groups. The multi-band {\eph} coupling constant and {\Tc}
are larger than the corresponding isotropic ones. With increasing
pressure, the isotropic and multi-band $\lambda$ and {\Tc} first
decrease and then increase, thus form a valley, and finally decrease
again. We also performed calculations using EPW, where the trend of
pressure dependent $\lambda$ and {\Tc} turns out to be rather similar
(see Appendix \ref{epwcal}). The obtained $\lambda$ are close to those
in Ref.~\onlinecite{Aoki1987,Rajagopalan1989,Nagara2010} but smaller
than those in Ref.~\onlinecite{Cohen2013}. The origin of this
difference is unclear, and may be related to an inherent difference of
the approaches. Overall, the trend as a function of pressure is not
consistent with the recent {\eph} coupling calculations using a
Wannier interpolation approach\cite{Cohen2013}, where a
large smearing parameter $\sigma$ (0.27 eV) is adopted which
significantly overestimates $T_c$ before the Lifshitz transitions.


From our theory, the calculated {\Tc} exhibits a valley formation similar
to Ref.~\onlinecite{Guo2016} depicted as down-triangles in
Fig. \ref{lambdatc}, as it is also consistent with the data in
Ref.~\onlinecite{Wittig1985} (green squares) in the range from
$P=5$~GPa to $P=25$~GPa. While our calculations can thus
explain the observed {\Tc} valley, the most prominent feature in
recent experiments, they do not explain the plateau seen in
Ref.~\onlinecite{Guo2016} beyond $P=32$~GPa. The {\Tc} valley can be
attributed to the Lifshitz transitions, which is also consistent with
the Hall measurements\cite{Guo2016}. The pressure for the {\Tc}
minimum in experiment and our theoretical calculations are at 17 GPa and 22
GPa, respectively. The difference is relatively small, and can be
decreased if we include correlation effects in the calculations. Two
state of the art methods, pseudopotential calculations combined with
Wannier function interpolation and full potential all-electron
calculations, arrive at the same result which can successfully explain
the {\Tc} valley formation.

\section{Discussion}
\begin{figure}[tb]
\includegraphics[width=\columnwidth]{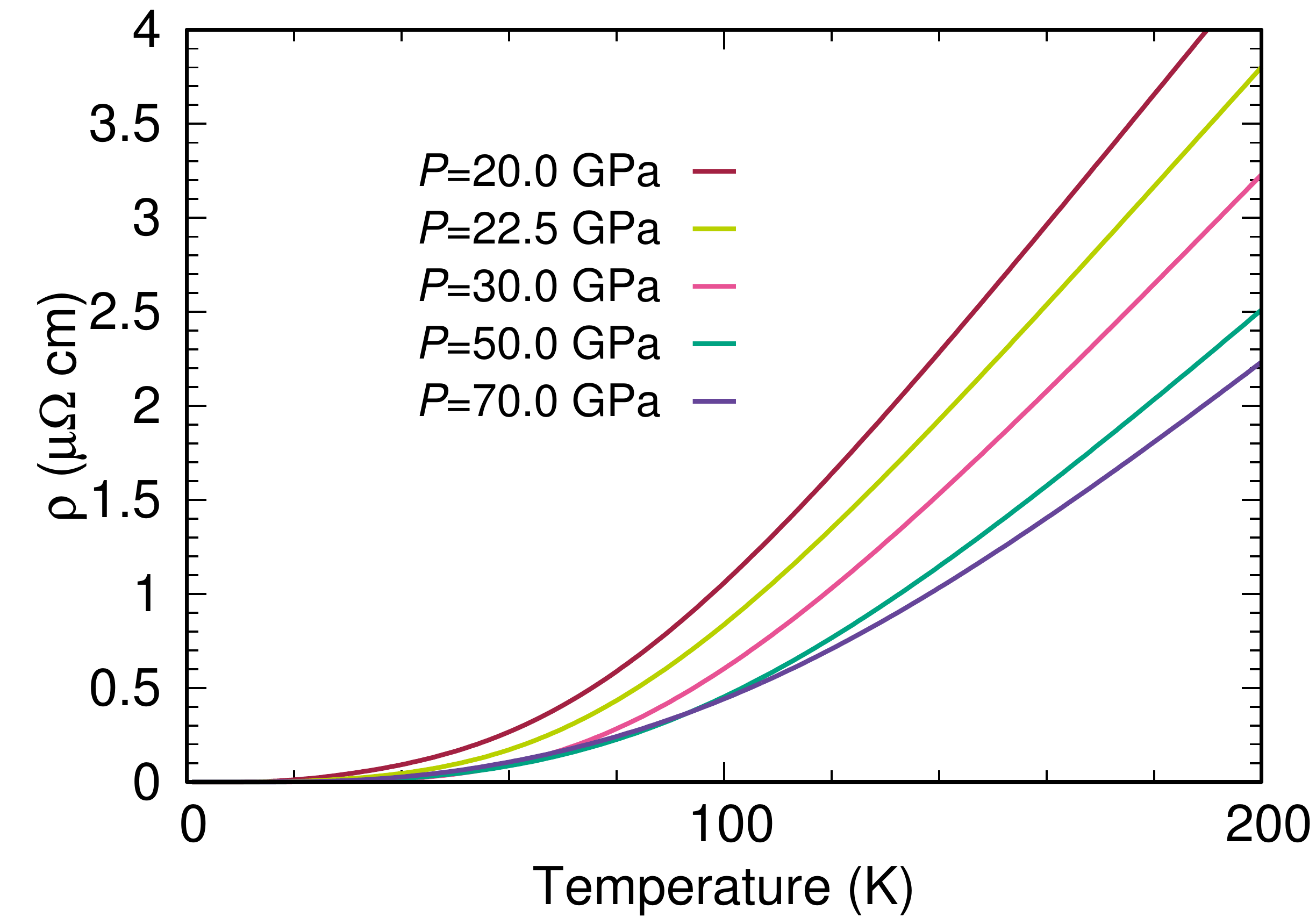}
\caption{(Color online) Temperature dependent
  resistivity for cubic phosphorus from {\eph}
  coupling.  \label{resistivity}}
\end{figure}

Above 5 GPa, when orthorhombic black phosphorus partially converts to
the rhombohedral phase, superconductivity appears. In the mixed phase
region, {\Tc} increases. At 12 GPa, the rhombohedral phase transforms
into the cubic phase, and {\Tc} shows a jump\cite{Guo2016}. Our
calculations indicate that the phase transition happens at 19 GPa (see Appendix \ref{HRphase}), which is higher than that in experiment. Moreover,
the couplings between $p$ orbitals along $[111]$ direction in the cubic
phase are uniform compared with those in the rhombohedral phase due to
its high-symmetry structure, which suppresses band splitting near the
Fermi level, and hence results in a higher $N(E_f)$. Therefore,
$N(E_f)$ shows a jump across the structural phase transitions, which
directly feeds into the value of {\Tc}.
Furthermore, we perform phonon calculations in both
phases at 22 GPa, where $d$ orbital bands are still not involved. The
phonon spectra are given in Appendix~\ref{HRphase}. Due to the change
in crystal symmetry at the phase transition, in the cubic phase there
are more degenerate phonon modes and the phonons soften a little compared with the rhombohedral phase. This leads to an
increased {\eph} coupling constant and {\Tc} shows a jump of about 1.3
K with $\mu^*=0.12$. It is consistent with experimental data in
Ref.~\onlinecite{Guo2016}, which shows that the structural phase transition
taking place at 12.4 GPa is accompanied by a jump in {\Tc}. At this low
pressure, the Lifshitz transition cannot occur according to our
calculations. Therefore, the {\Tc} jump across the phase transition
should be mainly attributed to the change in crystal symmetry rather
than the Lifshitz transition.

In the normal state, simple cubic phosphorus exhibits a quadratic
temperature dependence of the resistance in the normal state, and
resistance decreases as pressure increases\cite{Guo2016}. The
resistance due to {\eph} coupling can be expressed as\cite{Ziman1960},
\begin{equation}
\rho(T)=\frac{4\pi m_e}{ne^2k_B T}\int^{\infty}_{0}d\omega \hbar \omega \alpha^2_{\rm tr}F(\omega)n(\omega,T)[1+n(\omega,T)]\,,
\end{equation}
where $n(\omega,T)$ is the Bose-Einstein distribution. The Eliashberg
transport coupling function is similarly defined as the Eliashberg
spectral function\cite{Ziman1960}. According to our calculations, the
shape of $\alpha^2_{\rm tr}F(\omega)$ is very similar to that of
$\alpha^2F(\omega)$, and with increasing pressure $\alpha^2_{\rm
  tr}F(\omega)$ exhibits similar behavior compared to
$\alpha^2F(\omega)$. It will lead to a decreased resistance with
increasing pressure, as shown in Fig.~\ref{resistivity}, which is
consistent with the experimental data both in its temperature and its
pressure dependence. This suggests that {\eph} coupling has a
significant contribution to the electrical resistance.

Our calculations can successfully explain the observed {\Tc} valley in
the simple cubic phase but not the {\Tc} plateau at high pressure. The
{\Tc} valley in experiment turns out to have an electronic origin, as
it is due to Lifshitz transitions. At high pressure, the phonon modes
harden rapidly, and the {\eph} coupling constant as well as {\Tc} decrease with
increasing pressure in our calculations, which is inconsistent with
the observed {\Tc} plateau as a rather high value. If we study how {\Tc} evolves in
experiment for higher pressure, one finds that {\Tc} remains almost
unchanged for an extended regime of varying pressure, while the hole
carrier density decreases for $P>30$~GPa. Such evidence is rather
intriguing, and hard to explain. Let us first note that systematic errors
in the actual pressure imposed on the crystal might in fact be bigger
for higher than for lower pressures. As a consequence, it potentially
cannot be excluded that the increased pressure is not homogeneously
imposed on the whole crystal, and that the plateau might be in part
attributed to that. Instead, starting from the hypothesis that we take
this precise shape of {\Tc} as reference, one possible explanation
could be that electronic correlations may contribute to
superconductivity in simple cubic phosphorus at high pressure. From
the analysis above, we know that electron pockets around the
$M$ points are mostly composed of $d$ orbitals, and appear when
$P>22$~GPa. The electronic correlation effects in $d$ orbitals are
expected to be stronger than those in $p$ and $s$ orbitals. As such, these
correlation effects should become more and more important at high
pressures, even though the overall bandwidth increases. Assuming that electron-electron interactions might
participate in the pairing mechanism on these Fermi surfaces, with
increasing pressure, the $\delta$ Fermi surfaces grow bigger, and the
carrier density of electrons with $d$ orbital character increases (see
Fig.~\ref{dos_pressure}), which could enhance the Cooper pairing. This
will further enlarge the pressure region where {\Tc} increases. If the
enhancement from correlation effects can make up for the {\Tc}
decrease due to phonon hardening, it may lead to such a {\Tc} plateau
at high pressure.  In Ref.~\onlinecite{Flores-Livas2017}, where
electronic interactions have been considered within the static random
phase approximation (RPA), one could not reproduce the
experimentally observed behavior of {\Tc} at high pressure. This
suggests that effects which are beyond the static RPA level need to be
taken into account. One possibility is that the frequency dependence
of the interaction is important, and that plasmons, which can
cooperate with phonons and significantly raise {\Tc}, provide an
explanation~\cite{Akashi2013}.

\section{Conclusions}

We have investigated the electronic structure and electron phonon coupling
for simple cubic phosphorus by performing first-principle calculations
within the FP-LAPW method. Our calculations show that with increasing
pressure, {\Tc} first decreases, then increases and so forms a valley,
and then decreases again. Before the Lifshitz transitions set in, the
phonon hardening and shrinking of the $\gamma$ Fermi surface result in
a decrease of {\Tc} with increasing pressure. After the Lifshitz
transitions, the appearance of $\delta$ Fermi surfaces with $3d$
orbital character generates strong electron-phonon inter-band
couplings in $\alpha\delta$ and $\beta\delta$ channels, and hence
increases {\Tc}. With further increase of pressure, phonon hardening
makes a dominant contribution to electron-phonon coupling, and {\Tc}
decreases again. Our study reveals that the intriguing {\Tc} valley,
as found in
experiment should be attributed to the Lifshitz transitions. We also find,
however, that the experimentally observed {\Tc} plateau at high
pressure is beyond the electron-phonon mechanism considered here. It
suggests that besides electron-phonon coupling, plasmonic
contributions along with electronic correlations may be relevant for
systems with $d$-orbital character at higher pressures, such as for
simple cubic black phosphorous.

\acknowledgments

We particularly thank Gianni Profeta for highly useful discussions of
the electron-phonon coupling for multi-band systems. We also thank
Paolo Barone, Lilia Boeri, Miao Gao, Carla Verdi, Kay Dewhurst,
Giorgio Sangiovanni, and Gang Li for illuminating discussions. We also
thank Liling Sun and Honghong Wang for helpful discussions on the
experimental aspects. We acknowledge the Gauss Centre for Supercomputing e.V. (www.gauss-centre.eu) for funding this project by providing computing time on the GCS Supercomputer SuperMUC at Leibniz Supercomputing
Centre (LRZ, www.lrz.de). This work has been supported by
ERC-StG-TOPOLECTRICS-336012, SFB 1170 TOCOTRONICS, and SPP 1666. The work at Princeton was supported by the Gordon and Betty Moore Foundation EPiQS initiative, grant GBMF-4412.

\appendix
\section{Band structures for simple cubic phosphorus with GGA+U and HSE}\label{RCphase}

\begin{figure}[tb]
\includegraphics[width=\columnwidth]{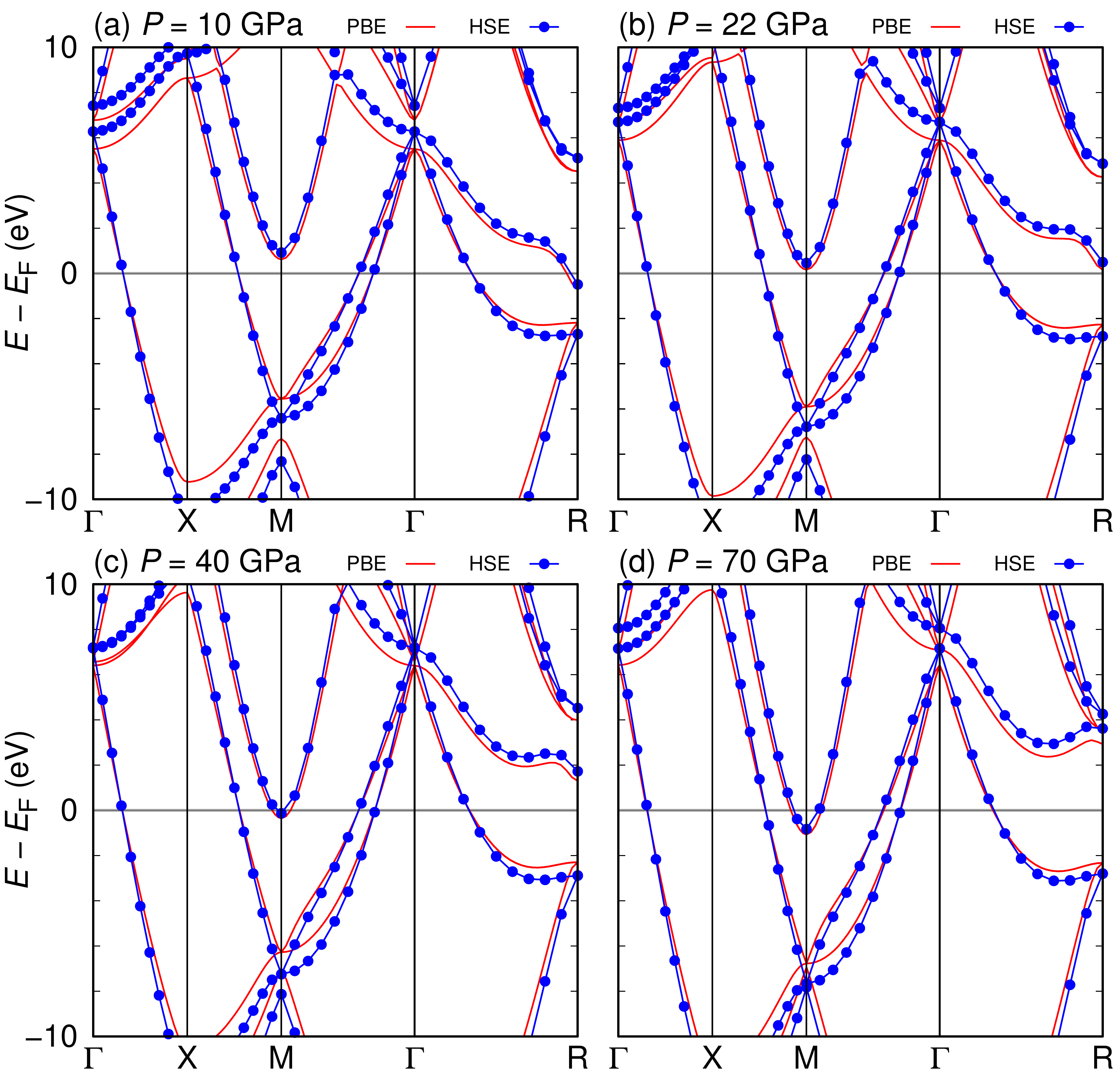}
\caption{(Color online) Band structures with GGA functional and
  hybrid functional (HSE06) for simple cubic phosphorous at four
  different external pressures. \label{bandhse}}
\end{figure}

According to our calculations, a band with $d$ orbital character will
cross the Fermi level at sufficiently high pressure. This suggests
that correlation effects may be important in simple cubic phosphorus,
and we check if they have an important effect on the electronic
structure. First, we perform calculations with a hybrid functional
(HSE06)\cite{Krukau2006}; the obtained band structures are shown in
Fig.~\ref{bandhse} in comparison with GGA bands. Compared to the bands
calculated with GGA functional, the bandwidth in HSE band increases,
and the most noticeable changes near the Fermi level are that the $s$
band at R and the $d$ band at M are are shifted to slightly higher
energies. The two Lifshitz transitions at the R and M points occur at
almost the same pressure in GGA calculations but happen at 20 GPa and
38 GPa in HSE calculations.

\begin{figure}[tb]
\includegraphics[width=\columnwidth]{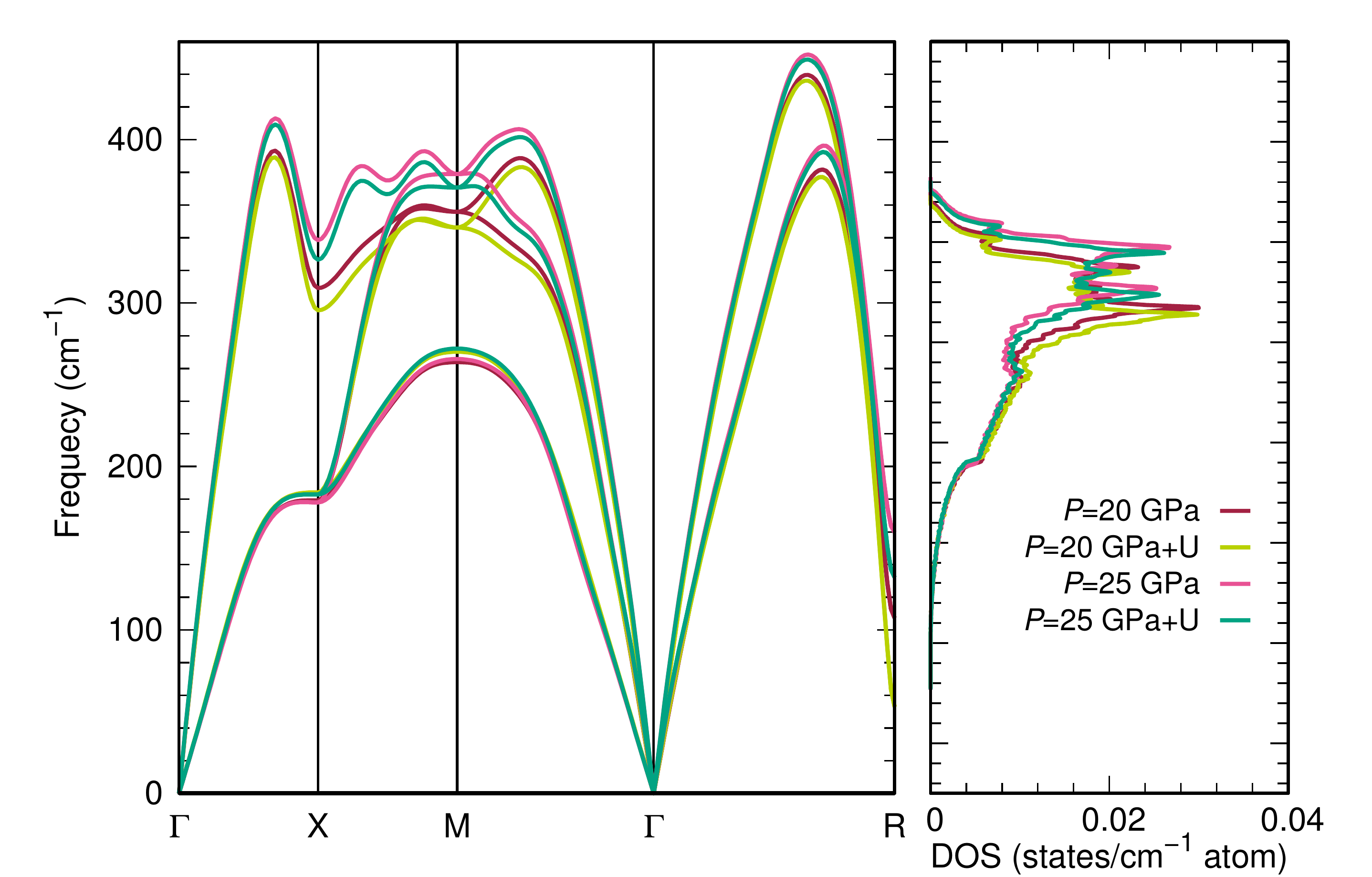}
\caption{(Color online) Phonon spectra and densities of states with
  GGA and GGA+U for simple cubic phosphorus at $P=$ 20 GPa and $P=$25
  GPa. \label{phggau}}
\end{figure}

We also tried to consider the local interaction by performing a GGA+U
calculation. The adopted $U$ and $J$ are 2.0 eV and 0.5 eV for $p$
orbital and $d$ orbital, respectively. The band structures are given in
Fig.~\ref{phggau}. No matter whether the additional $U$ interaction is
on $p$ or $d$ orbitals, the obtained band structures are almost the
same as those with GGA functional. Therefore, GGA+U has little effect
on the band structures of simple cubic phosphorus. Furthermore, we
calculate the phonon spectra for phosphorus at 20 GPa and 25 GPa, and
the phonon dispersions are shown in Fig.~\ref{phggau}. The phonon
modes in GGA+U calculations soften a little bit compared with those in
GGA calculations. This phonon softening will increase the {\eph}
coupling but cannot change the general negative trend of the pressure-dependent {\eph}
coupling constant.

\begin{figure}[tb]
\includegraphics[width=\columnwidth]{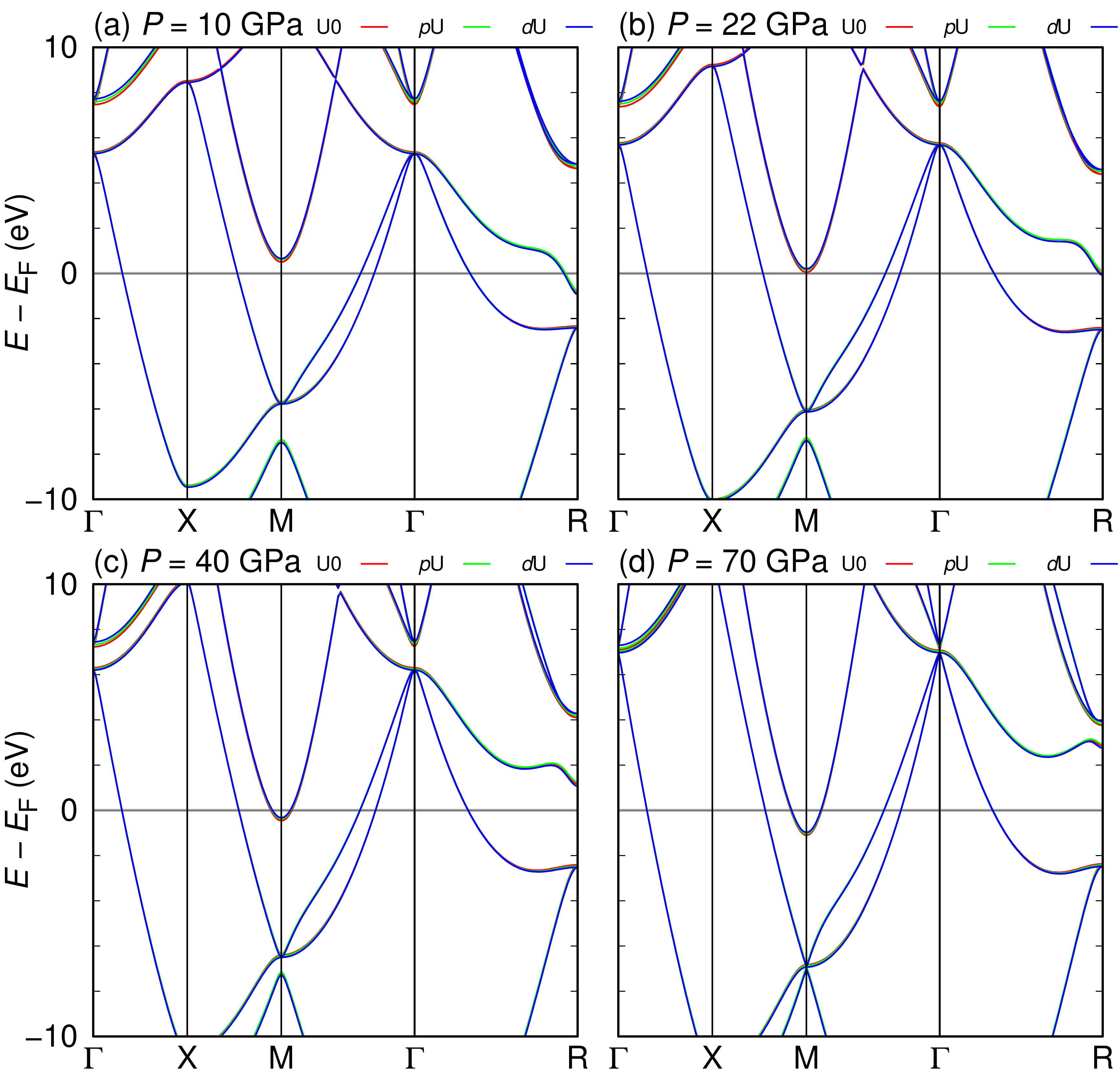}
\caption{(Color online) The band structures with GGA+U for simple
  cubic phosphorous at four different external pressures. $U0$
  represent GGA calculations, $pU$ and $dU$ represent that in GGA+U
  calculations the additional U is on $p$ orbital and $d$ orbital,
  respectively.  \label{bandggau}}
\end{figure}

\begin{figure}[tb]
\includegraphics[width=\columnwidth]{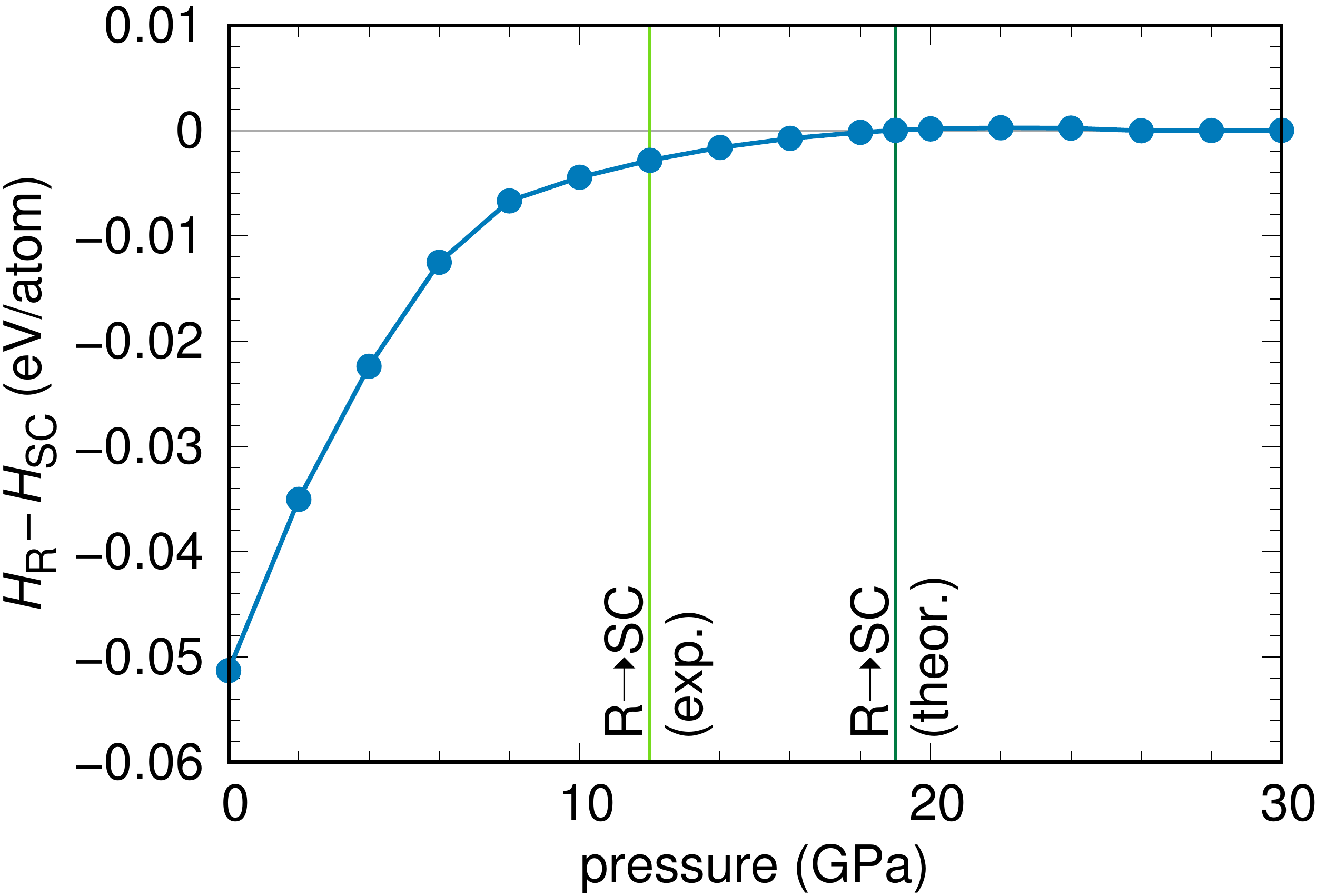}
\caption{(Color online) Enthalpy for rhombohedral phosphorus relative
  to the simple cubic phase as a function of pressure. \label{Renthalpy}}
\end{figure}

\begin{figure}[tb]
\includegraphics[width=\columnwidth]{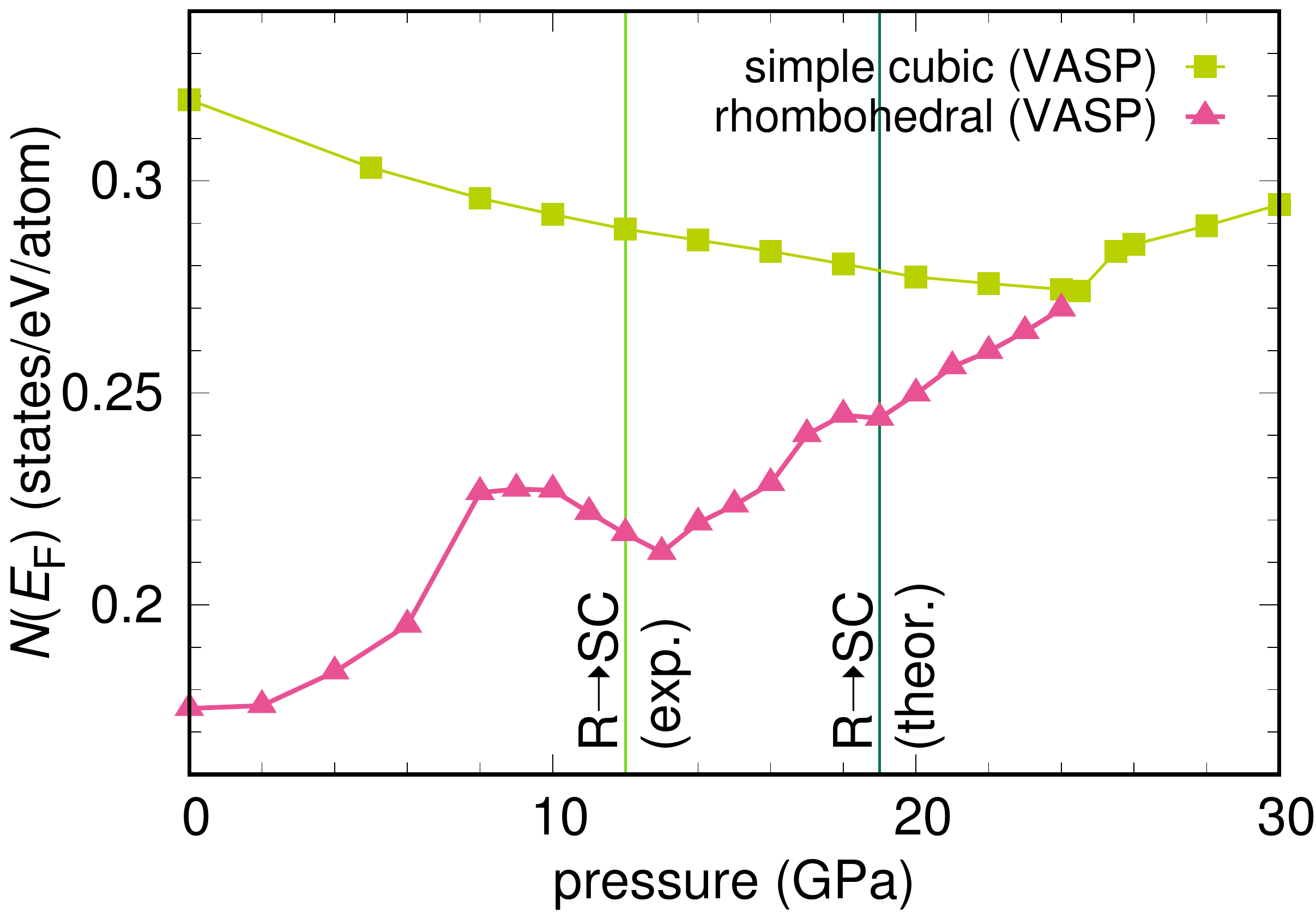}
\caption{(Color online) Density of states at the Fermi level for the
  simple cubic and rhombohedral phases as a function of
  pressure. \label{RCdos}}
\end{figure}

\section{Electron-phonon coupling constant and {\Tc} from EPW} \label{epwcal}

\begin{figure}[tb]
\includegraphics[width=\columnwidth]{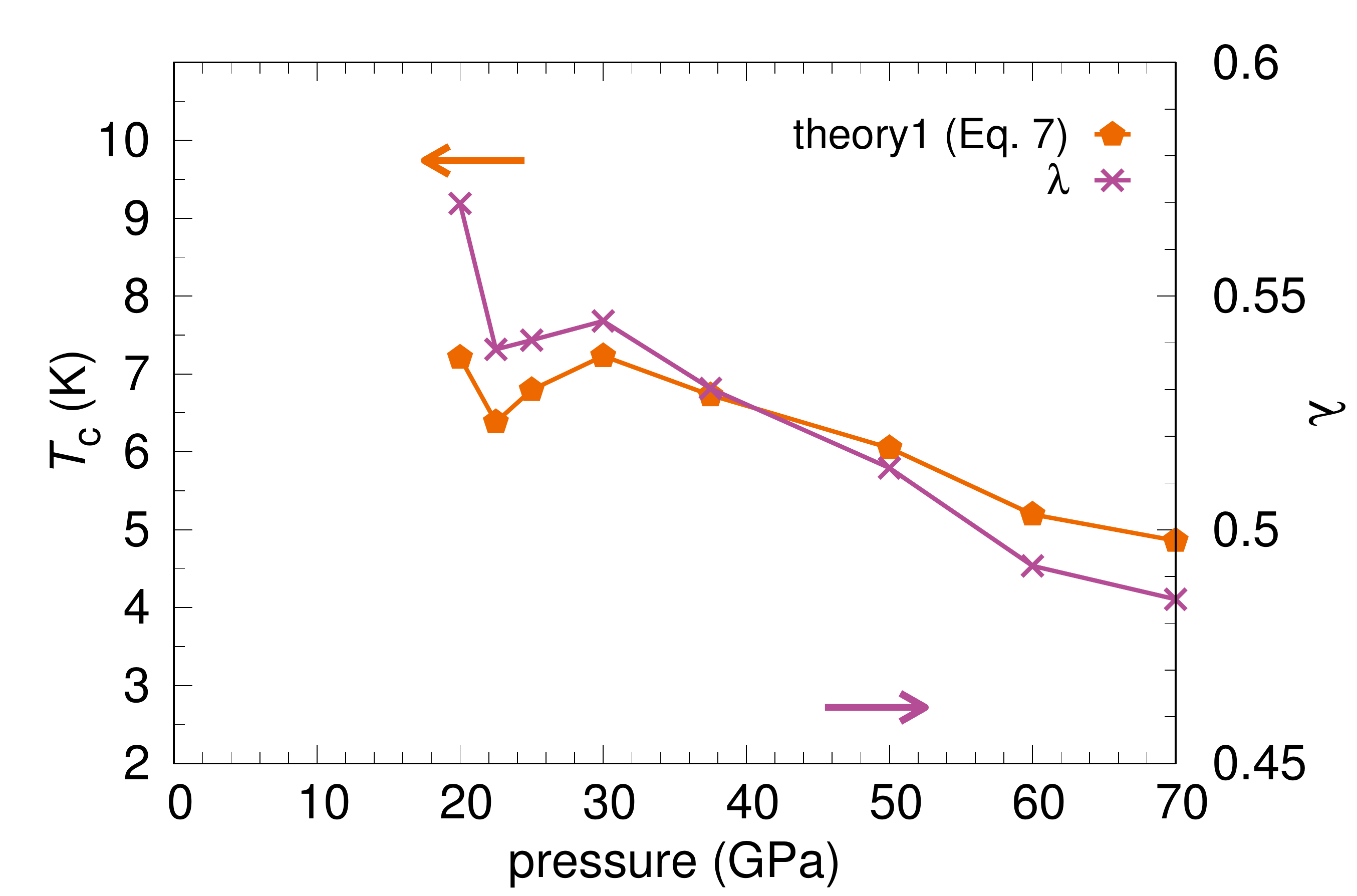}
\caption{(Color online) Superconducting transition temperature {\Tc} and {\eph} coupling constant $\lambda$ as a function
  of pressure in theoretical calculations using EPW. The adopted
  $\mu^{*}$ is 0.1.  \label{tcepw}}
\end{figure}

We also perform electron-phonon coupling calculations using EPW. To
obtain the nine Wannier functions, we use the Bloch functions on an
$8\times8\times8$ $k$ mesh. Phonon dispersions are calculated on an
$8\times8\times8$ $q$-mesh using density functional perturbation
theory. For the {\eph} calculations, a very fine
$128\times128\times128$ $k$ mesh and a $24\times24\times24$ $q$ mesh
are used, and the $\delta$ function in the phonon linewidth is
approximated by Gaussian functions with $\sigma=0.08$ eV.  After
obtaining good Wannier functions, which can be used to reproduce the
first-principle band structures and phonon dispersions with high
accuracy, we perform electron-phonon calculations. The obtained
electron-phonon coupling constant and {\Tc} are shown in
Fig.~\ref{tcepw}. Similar to the results from elk, both $\lambda$ and
{\Tc} first decrease then increase, forming a valley, and decrease
again with increasing pressure. Therefore, both elk and EPW
calculations confirm that there is a {\Tc} valley in the
superconducting {\Tc} versus pressure phase diagram, which can be
attributed to the Lifshitz transitions. The results are consistent
with those from elk calculations.

\section{Density of states evolution from rhombohedral to cubic
  phosphorus} \label{HRphase}

While we have focused on simple cubic phosphorous in this work as it
is the phase with the highest superconducting transition temperature,
it is also interesting how the transition to the high {\Tc} phase
occurs.  We first investigate the stability of rhombohedral phosphorus
relative to simple cubic phosphorus by calculating enthalpies as a
function of external pressure; the difference between enthalpies in
simple cubic and rhombohedral phases is shown in
Fig.~\ref{Renthalpy}. Below 9~Gpa, the dominant structure is
orthorhombic but here we focus on the cubic and rhombohedral phases
above 10~GPa. From our calculations, the enthalpies of the two phase
are very close above 10~GPa and the rhombohedral to simple cubic phase
transition occurs at 19~GPa, which is consistent with previous
calculations\cite{Cohen2013}. This transition pressure is higher than
the 12~GPa observed in experiment\cite{Guo2016}. Fig.~\ref{RCdos}
shows the DOS at the Fermi level for simple cubic and rhombohedral
phases as a function of pressure. At the phase transition, the density
of states at the Fermi level $N(E_F)$ jumps up sharply (both at the
experimental and at the theoretical transition pressures). As {\Tc} is
often proportional to $e^{-\frac{1}{VN(E_F)}}$, the jump of $N(E_F)$
can lead to a jump in {\Tc}, which explains the observed {\Tc} jump
across the phase transition in experiments\cite{Guo2016}.  Moreover,
the {\Tc} increase from 20 to 25~GPa in
Ref.~\onlinecite{Flores-Livas2017} should partially be due to the
$N(E_F)$ increase. However, X-ray diffraction measurements show that
the cubic phase is pure above 13.44 GPa\cite{Guo2016}. The adopted
structure is rhombohedral in the pressure range from 10 to 20~GPa in
calculations\cite{Flores-Livas2017}, which is not consistent with the
experimental data from Guo \emph{et al.}\cite{Guo2016}. In the simple
cubic phase, our conclusion is consistent with that of
Ref.~\onlinecite{Flores-Livas2017} that {\Tc} will decrease with
increasing pressure (at high pressure).

\begin{figure}[t]
\includegraphics[width=\columnwidth]{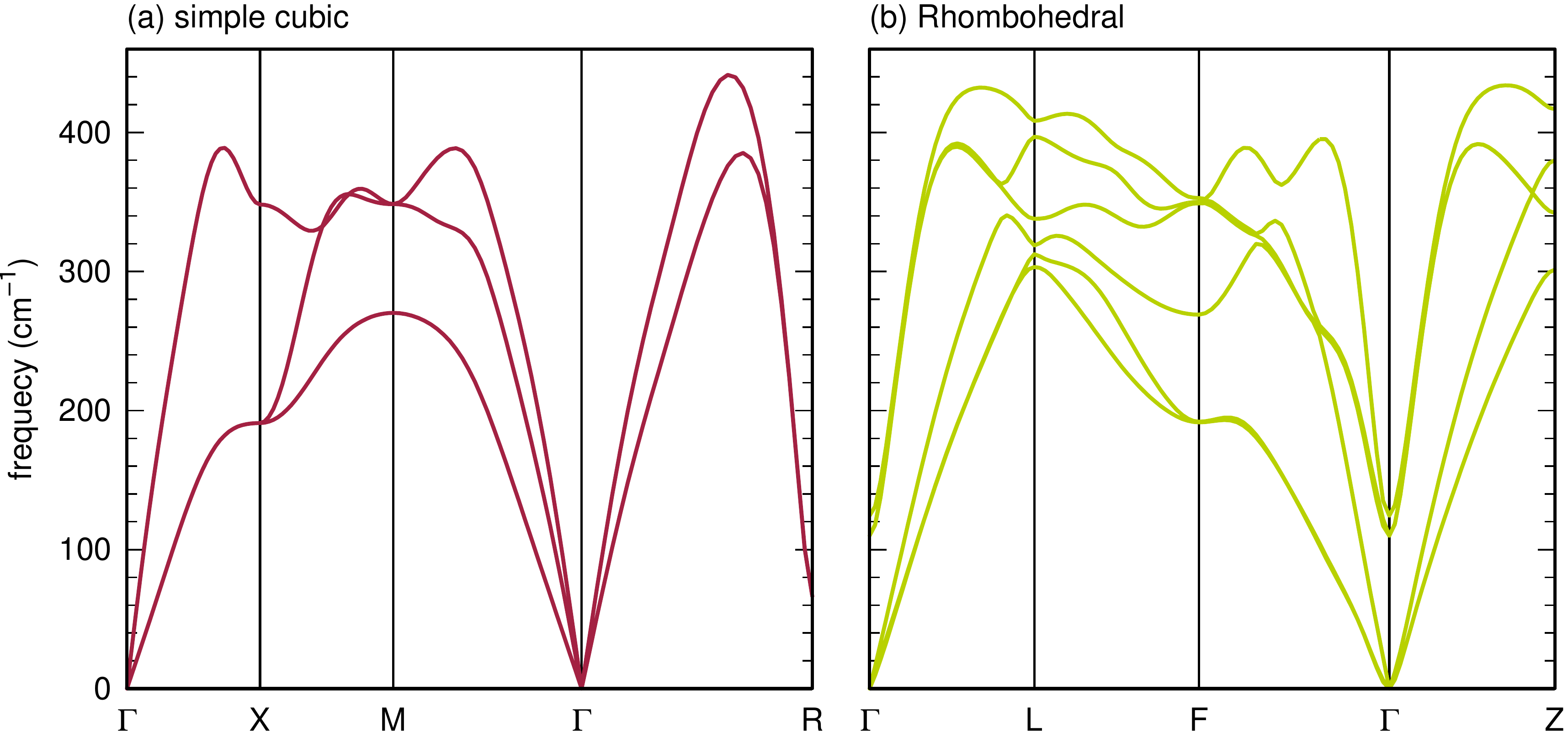}
\caption{(Color online) Phonon spectra for the simple cubic and
  rhombohedral phases at 22 GPa. \label{RCph}}
\end{figure}

\begin{figure}[tb]
\includegraphics[width=\columnwidth]{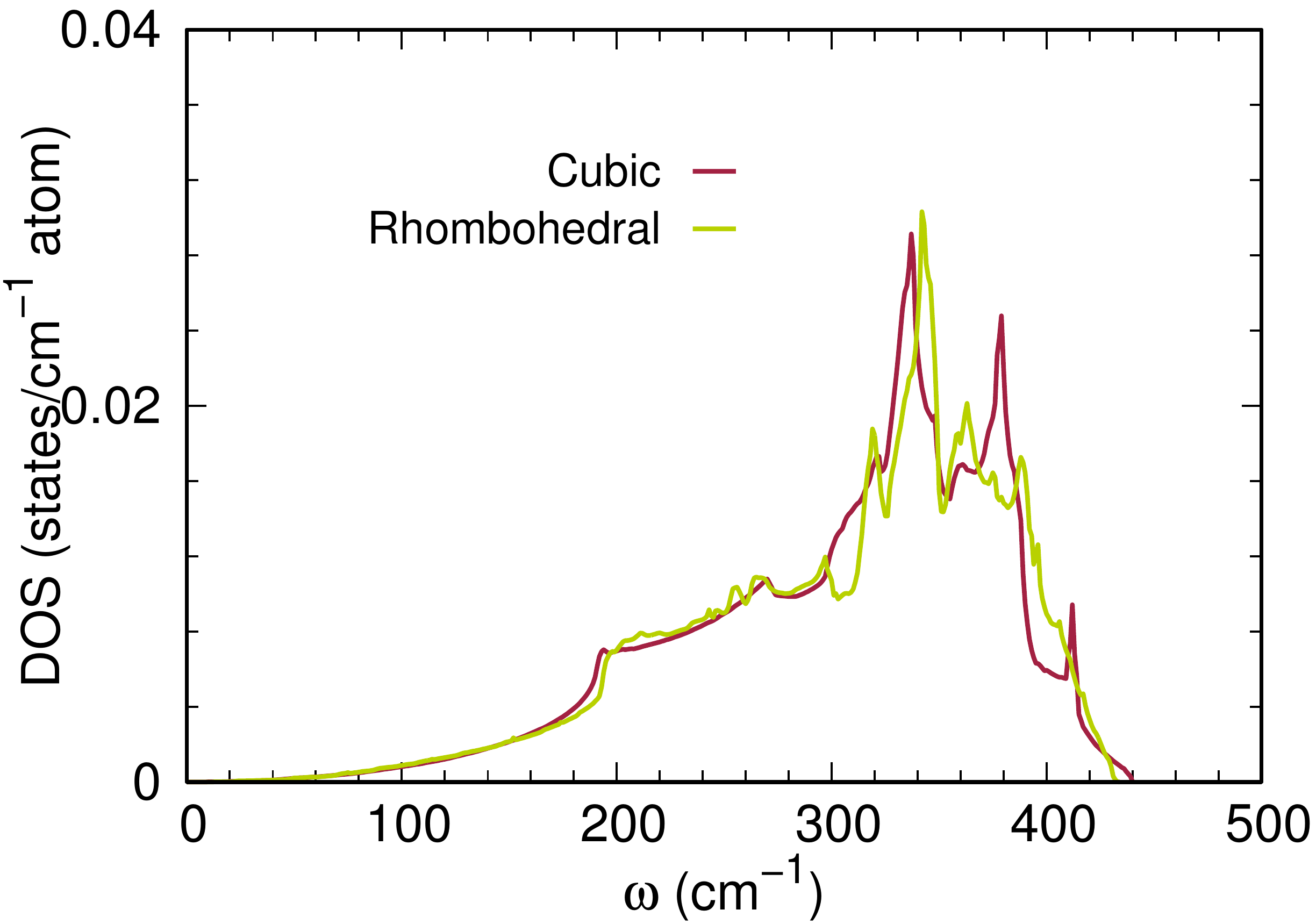}
\caption{(Color online) Phonon density of states for the simple cubic
  and rhombohedral phases at 22 GPa. \label{RCphdos}}
\end{figure}

To estimate if {\Tc} will jump across the phase transition, we perform
calculations for phosphorus in simple cubic and rhombohedral (R)
phases. As there are unstable phonon modes in the cubic phase at low
pressure, we perform the calculations at 22 GPa, near the theoretical
phase transition point, where $N(E_F)$ still shows a jump. It can
represent the general behavior of {\Tc} during the phase
transition. At this pressure, the $d$ band is close to the Fermi level
but does not cross it in either phase. The phonon spectra and DOS are
shown in Fig.~\ref{RCph} and \ref{RCphdos} and their DOS are very
similar. In the R phase the phonon modes harden a little and optical
modes show a splitting compared with the cubic phase due to additional
coupling in the lower-symmetry crystal. If we assume the same
effective interaction in the two phases, the {\eph} coupling constant
is proportional to $N(E_F)$. Using the McMillan equation, we estimate
that the {\Tc} jump is about 1.3 K with with $\mu^*=0.12$. At the
lower pressure, the phonon hardening in the R phase will be
larger. The {\Tc} jump will be bigger at the experimental phase
transition point because the atomic distortion in the R phase is
larger than that in the present calculations. Therefore, when the R
phase transforms into the simple cubic phase, {\Tc} will show a jump
and it can be mainly attributed to the change of crystal symmetry
(affecting the electronic structure and phonon modes) rather than to
the Lifshitz transition as suggested in
Ref.~\onlinecite{Flores-Livas2017}. This is also supported by
experimental data which shows that the phase transition (rhombohedral
to simple cubic) happens at 12 GPa\cite{Guo2016} accompanied by a jump
in {\Tc} where the Lifshitz transition cannot occur.

\end{document}